\documentclass[11pt,a4paper]{article}
\usepackage[margin=25mm]{geometry}
\usepackage[T1]{fontenc}
\usepackage{lmodern}  %
\usepackage{amsmath,amssymb}
\usepackage{graphicx}
\graphicspath{{figures/}}
\usepackage{booktabs,array}
\usepackage{caption,subcaption}
\usepackage{placeins}  %
\usepackage[numbers,sort&compress]{natbib}
\usepackage{xcolor}
\usepackage[hidelinks]{hyperref}
\usepackage[capitalise,nameinlink]{cleveref}

\newcommand{\Vs}{V^{*}}
\newcommand{\Vf}{V^{*}_{f}}
\newcommand{\rr}{\raggedright\arraybackslash}  %

\newcommand{\bibdoi}[1]{(doi: \href{https://doi.org/#1}{\nolinkurl{#1}})}

\title{How the elastic axis folds the transonic flutter boundary: shock energy transfer
and conformally calibrated active learning}

\author{Ashray Saxena\textsuperscript{*} and Sharad Shrivastava\\[4pt]
\small Department of Mechanical Engineering, Birla Institute of Technology and Science,\\
\small Pilani, Vidya Vihar, Pilani, Rajasthan 333031, India\\[4pt]
\small \textsuperscript{*}Corresponding author: f20231043@pilani.bits-pilani.ac.in,
ORCID \href{https://orcid.org/0009-0001-8570-9160}{0009-0001-8570-9160}\\
\small Sharad Shrivastava: sharad\_shrivastava@pilani.bits-pilani.ac.in}
\date{\small Accepted for publication in \emph{Transport Phenomena} (De Gruyter Brill),
\href{https://doi.org/10.1515/tp-2026-0137}{doi:10.1515/tp-2026-0137}}

\begin{document}
\maketitle

\begin{abstract}
Isogai's case A, a pitch--plunge aerofoil whose elastic axis lies half a chord ahead of the
leading edge, has a folded transonic flutter boundary in inviscid flow, with several flutter
speeds at one Mach number. We map this boundary and its elastic-axis family with coupled Euler
simulations, a Gaussian-process surrogate of the damping trained by batch
active learning, and distribution-free split conformal prediction, which bounds the marginal error rate of stability calls.
Three failure modes of damping labels
are invisible in the residuals, and the half-degree mean incidence
left by an impulsive start at $1^\circ$ freestream
incidence raises the flutter speed at the dip by 2.8\%, more than any numerical setting.
With 80 simulations the learned case A boundary labels 63 of 64
independent runs correctly, and at most 1.8\% of calibrated calls on held-out runs are wrong. Over Mach number,
elastic axis and flutter speed index, 193 simulations show
that, for the case~A structure about the axis, the inviscid fold exists only for
axes more than about 0.4 chord ahead of the leading edge, on two meshes.
Chordwise energy budgets, closed three ways, explain both branches. On the lower branch the
falling reduced frequency, at unchanged shock phase, raises the shock input and lowers the
upstream damping similarly. On the upper branch the
shock is pinned at the trailing edge; moving the axis aft brings the
flutter-mode pivot towards the section, so the upstream damping falls faster than the
residual shock input and the fold closes.
\end{abstract}

\noindent\textbf{Keywords:} aeroelasticity; computational fluid dynamics; surrogate
modelling; uncertainty quantification; unsteady aerodynamics

\section{Introduction}
\label{sec:intro}
Flutter is a self-excited oscillation in which the flow does net positive work on an
elastic structure over each cycle \cite{dowell2001}. At transonic speeds the flutter
speed falls steeply over a narrow band of Mach numbers, the transonic dip, because the
shock waves on the surface move with the structure and lag behind it
\cite{tijdeman1980,bendiksen2011}. Energy arguments have explained the dip since Ashley's
shock-doublet model \cite{ashley1980}: the work of the moving shock has been separated
from that of the other flow regions \cite{bendiksen1992}, computed for Isogai's case~A
\cite{pradeepa2012,karnick2017} and resolved along the chord of a wing section
\cite{kuzmina2020}; a complementary view couples the structure to a stable fluid mode
\cite{gao2020}. Beyond the boundary, inviscid aerodynamic nonlinearity associated with the
moving shocks can limit the motion to limit cycles \cite{kousen1994,thomas2002,dowell2003}. Linear aerodynamics
contains none of this, so transonic flutter boundaries are computed with coupled
computational fluid dynamics (CFD). Because each coupled simulation is expensive,
the flutter point has also been located by solving directly for the Hopf bifurcation
\cite{badcock2004}, by searching the parameter space with a hierarchy of aerodynamic models
\cite{timme2011}, and with reduced-order and data-driven models of the unsteady
aerodynamics \cite{lucia2004,li2019,kou2021}.

Isogai's case~A \cite{isogai1979}, a pitch--plunge section with a far-forward elastic axis
and a large static unbalance, is the standard test of such computations. In inviscid flow
its boundary is folded: between Mach numbers of about 0.85 and 0.9 several flutter speeds
coexist \cite{alonso1994,liu2001,marques2020}; viscosity, included in thin-layer and
Reynolds-averaged Navier--Stokes analyses of such sections \cite{prananta1998,karnick2017},
removes them but keeps the dip \cite{karnick2017}. With small-disturbance aerodynamics \citet{isogai1981} traced the dip
to the lagging shock acting on a mode that pivots far ahead of the leading edge, and the
sixfold rise of the flutter speed at $M_\infty=0.90$ to the shock reaching the trailing
edge. The elastic axis has been varied with the Euler equations at the single Mach number
0.80 \cite{pradeepa2012}, and harmonic-balance Euler studies found several flutter speeds
at one Mach number for large mass ratios with the elastic axis at $a=-0.6$
\cite{kholodar2003}, and mapped the flutter speed over Mach number and mass ratio
\cite{kholodar2004}. How the fold emerges as the elastic axis moves forward towards
Isogai's configuration, and where it closes, has not, to our knowledge, been mapped.

Each point of such a map is a coupled transient whose damping must be identified.
Sampling an expensive simulator where a Gaussian process is most uncertain about a
level set was developed for contour estimation and structural reliability
\cite{ranjan2008,bichon2008,echard2011}, and the uncertainty of flutter predictions has long
been a concern of aeroelasticity \cite{pettit2004}.
\citet{marques2020} learned the case~A boundary with a Gaussian-process surrogate sampled
where its zero contour is most uncertain, at a fraction of the cost of bisection; adaptive
surrogates have since given probabilistic boundaries under structural uncertainty
\cite{lu2025}, with batch sampling for parallel computers \cite{lu2026}. Conformal
prediction, which gives prediction intervals with finite-sample coverage under
exchangeability alone \cite{vovk2005,lei2018,angelopoulos2023}, has been combined with
active-learning kriging in structural reliability \cite{jaber2026}, but not, to our
knowledge, for flutter. Two questions remain open: are
the damping labels reliable, when their sign can be wrong for short or coarsely resolved
records \cite{jacobson2019,mendez2026}, and how far can a learned boundary be trusted
between simulations?

We answer both and use the answers to map how the elastic axis shapes the boundary. To
our knowledge the following are new.
(i) Three silent failure modes of time-domain damping labels, each with a check, and the
effect of the small mean incidence left by the usual $1^\circ$ start, which exceeds every
numerical error and is consistent with the
difference between the two published values at the bottom of the dip.
(ii) Stability calls with a distribution-free bound on their marginal error
rate \cite{vovk2005,lei2018}, from
split conformal calibration on independent random simulations, for case~A and for the
elastic-axis family, tested on held-out runs.
(iii) The calibrated boundary over Mach number, elastic axis and flutter speed index:
in inviscid flow and for the case~A structure about the axis, multiple flutter points exist only for an elastic axis more than
about 0.4 chord ahead of
the leading edge, on two meshes. Matched simulations at one flow condition show why: as the
axis moves aft, the pivot of the flutter mode approaches the section and the damping of the
supersonic surface falls faster than the input of the shock pinned at the trailing edge.
(iv) A chordwise energy budget of both branches, closed against the solver's loads and the
growth rate, with the work of the shock separated from that of the smooth pressure
field. The lower branch is crossed through the reduced frequency at unchanged shock
phase and mode shape, which changes the shock input and the upstream damping by
similar amounts; on the upper branch, where the shock reaches the trailing edge
\cite{isogai1981}, the collapse of the shock input is partly offset by a loss of upstream
damping.

\Cref{sec:formulation} states the problem. \Cref{sec:hifi} describes the coupled
solver, the identification of the damping and its verification, including the failure
modes and the effect of the start. \Cref{sec:learning} describes the surrogate, the active
learning and the conformal calibration. \Cref{sec:results} gives the case~A boundary, its
energy transfer, the elastic-axis family and the mechanisms of the two branches, and
\cref{sec:discussion} discusses them; \cref{app:verification} collects the solver settings
and the sensitivity studies.

\section{Problem formulation}
\label{sec:formulation}

\begin{figure}[t]
\centering
\includegraphics[width=\textwidth]{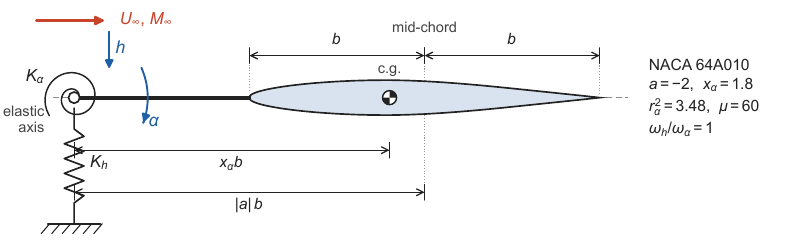}
\caption{The typical section of Isogai's case~A, drawn to scale. The elastic axis lies
$|a|b=c$ ahead of mid-chord, half a chord ahead of the leading edge, and the centre of
gravity $x_\alpha b$ aft of it.}
\label{fig:schematic}
\end{figure}

A rigid aerofoil of chord $c=2b$ is supported by a plunge spring and a torsion spring
about an elastic axis $ab$ aft of mid-chord (\cref{fig:schematic}), with plunge $h$
positive downward and pitch $\alpha$ positive nose-up. With time scaled by the uncoupled
pitch frequency $\omega_\alpha$ and $\boldsymbol{\eta}=[h/b,\ \alpha]^{\mathsf T}$,
\begin{equation}
  \begin{bmatrix} 1 & x_\alpha \\ x_\alpha & r_\alpha^2 \end{bmatrix}\ddot{\boldsymbol{\eta}}
  + \begin{bmatrix} (\omega_h/\omega_\alpha)^2 & 0 \\ 0 & r_\alpha^2 \end{bmatrix}\boldsymbol{\eta}
  = \frac{\Vs{}^2}{\pi}\begin{bmatrix} -c_l \\ 2c_m \end{bmatrix},
  \qquad
  \Vs=\frac{U_\infty}{b\,\omega_\alpha\sqrt{\mu}},
  \label{eq:eom}
\end{equation}
where $x_\alpha$ and $r_\alpha$ are the static unbalance and radius of gyration about the
elastic axis in semichords, $\mu=m/(\pi\rho_\infty b^2)$ is the mass ratio and $c_m$ is
taken about the elastic axis \cite{alonso1994,marques2020}; structural damping is zero.
Case~A \cite{isogai1979} is the NACA~64A010 section with $a=-2$, $x_\alpha=1.8$,
$r_\alpha^2=3.48$, $\omega_h/\omega_\alpha=1$ and $\mu=60$. We vary $a$ alone and keep $x_\alpha$, $r_\alpha$,
$\mu$ and $\omega_h/\omega_\alpha$ at their case~A values, so that the matrices of
\eqref{eq:eom} are the same for every member of the family. The family is therefore one of
structural configurations, not the displacement of the support under an unchanged section.
Since $x_\alpha$ and $r_\alpha$ are measured from the elastic axis, the centre of gravity
moves with it and stays $x_\alpha b=0.9c$ aft of it, at $x_{\mathrm{cg}}=c(1+a+x_\alpha)/2$
from the leading edge: from $0.15c$ at $a=-2.5$ through $0.4c$ for case~A to $0.9c$ at
$a=-1$. The radius of gyration about the centre of gravity, $(r_\alpha^2-x_\alpha^2)^{1/2}b
=0.49b$, does not change. What changes is the aerodynamic lever arm: the moment $c_m$ about
the axis, and the motion of the surface for a given pitch. A section of fixed mass
distribution on a moving support would change $x_\alpha$ and $r_\alpha$ with $a$; it is
compared with the family at the closure of the fold in \cref{sec:closure}.

A small disturbance decays or grows as a sum of modes $e^{(\sigma_j+i\omega_j)t}$. The
least-stable
mode is the one with the largest $\sigma_j$, and its damping coefficient
$\gamma=\sigma_j/\omega_j$, by which the amplitude changes by the factor $e^{2\pi\gamma}$ per
cycle, defines the
flutter boundary as the zero level set $\mathcal{Z}=\{\mathbf{x}:\gamma(\mathbf{x})=0\}$ over
$\mathbf{x}=(M_\infty,a,\Vs)$. Because $\mathcal{Z}$ is folded near the dip of case~A
\cite{alonso1994,karnick2017}, it is learned as a contour, not as a function
$\Vf(M_\infty,a)$.

\section{High-fidelity model and verification}
\label{sec:hifi}

\subsection{Coupled Euler solver and meshes}
The unsteady Euler equations are solved with SU2 8.5.0 \cite{economon2016} on deforming
meshes: Jameson--Schmidt--Turkel fluxes (dissipation coefficients 0.5 and 0.02),
Green--Gauss gradients and second-order dual time stepping, with implicit pseudo-time
iterations (CFL number 50, FGMRES with ILU) until the density residual falls below
$10^{-6}$, at most 80 per step. SU2's typical-section model \cite{alonso1994} re-solves
\eqref{eq:eom} with the current loads every five inner iterations (strong coupling); every
step needed at least 16 inner iterations, so the structure was updated at least three times
per step, which excludes the missed-update defect noted by \citet{mendez2026}. The section
pitches about the elastic axis, $x_{\mathrm{ea}}=c(1+a)/2$ from the leading edge, and the
volume mesh follows it through SU2's volumetric movement (linear elasticity with
inverse-volume stiffness); the configuration keys of the coupling and the mesh motion
are listed in \cref{tab:su2}. With $\omega_\alpha=100$~rad\,s$^{-1}$, $\Vs$ is set through
$U_\infty$ and the Mach number through the freestream temperature. Each run is released
from the steady solution at $1^\circ$ incidence into a zero-incidence freestream, so the
section oscillates about the symmetric state (\cref{sec:incidence} shows why this matters),
and is advanced for 12 pitch periods with 128 time steps per period.

\begin{figure}[t]
\centering
\includegraphics[width=\textwidth]{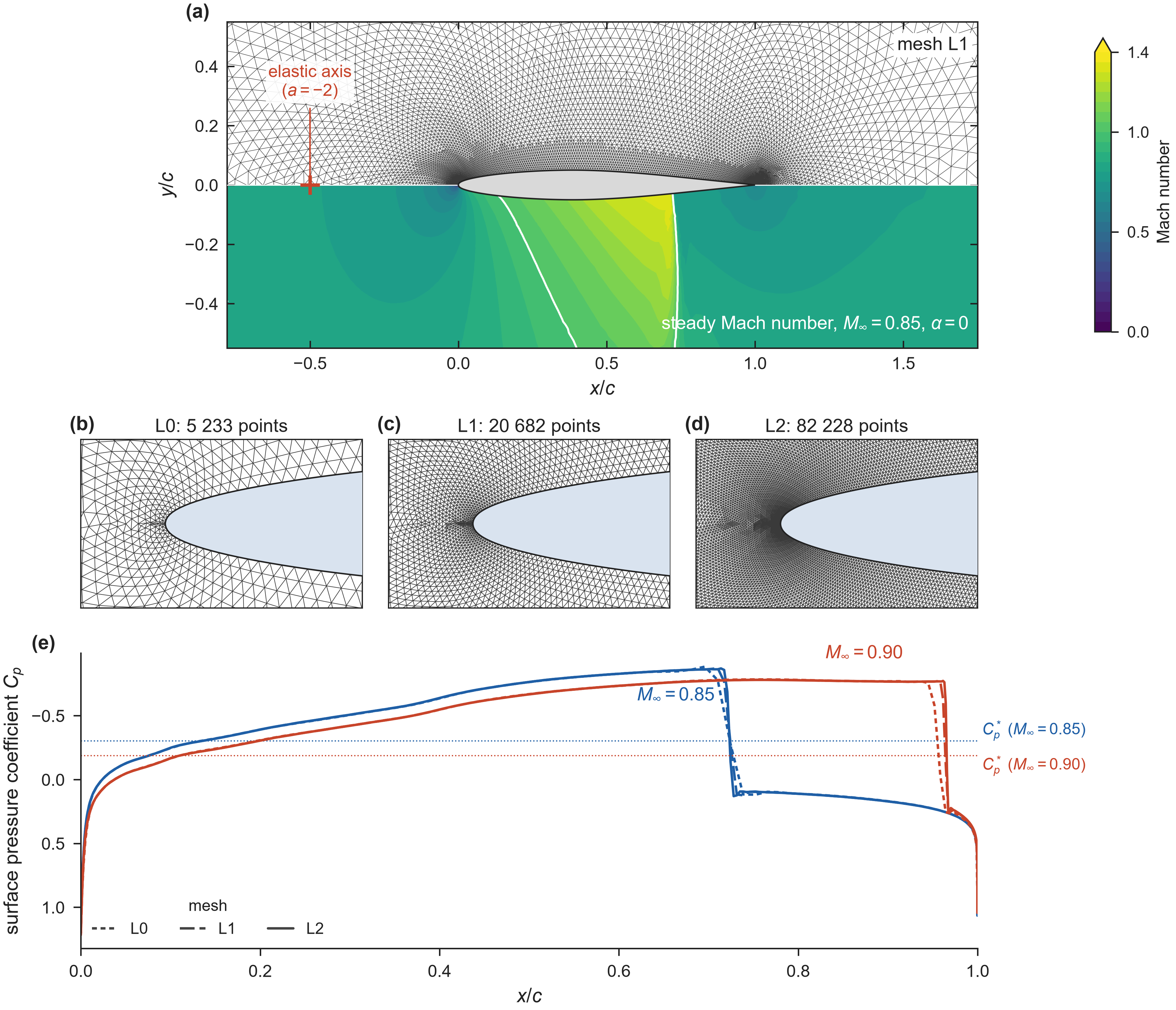}
\caption{Meshes and steady flow. (a) Near field of mesh L1 (upper half) and steady Mach
number at $M_\infty=0.85$, zero incidence (lower half; sonic line white, cross: elastic axis
of case~A). (b--d) Leading-edge region of meshes L0--L2. (e) Upper-surface pressure
coefficient at $M_\infty=0.85$ and 0.90 on the three meshes.}
\label{fig:meshes}
\end{figure}

The section is a cubic spline in $\sqrt{x/c}$ through the tabulated NACA~64A010
ordinates\footnote{UIUC airfoil coordinates database,
\url{https://m-selig.ae.illinois.edu/ads/coord_database.html}, file \texttt{naca64a010.dat}.},
with continuous slope and curvature. Straight-line interpolation of the table, an easy
shortcut, leaves slope jumps of up to $1^\circ$, each of which puts a visible step into the
transonic surface pressure. Three nested triangular meshes with the far field at $20c$ are
used (\cref{fig:meshes}): L0 (5\,233 points, 200 on the aerofoil), made from the inviscid
NACA~0012 mesh of the SU2 tutorials by moving its wall nodes onto the section, and L1
(20\,682) and L2 (82\,228) by subdividing every triangle into four (smallest angle
$17.0^\circ$, largest aspect ratio 3.4). Steady solutions at zero incidence give
$|c_l|\le0.0033$; the drag converges at an observed order of 2.5 at $M_\infty=0.90$ and
agrees within 0.4\% at 0.85 (\cref{tab:steady}).

\begin{table}[t]
\centering
\caption{Steady rigid NACA~64A010 at zero incidence: drag coefficient on the three
meshes, observed order $p$ and Richardson extrapolation $c_{d,\infty}$.}
\label{tab:steady}
\begin{tabular}{lccccc}
\toprule
$M_\infty$ & L0 & L1 & L2 & $p$ & $c_{d,\infty}$ \\
\midrule
0.85 & 0.023233 & 0.023322 & 0.023233 & -- & -- \\
0.90 & 0.070917 & 0.072847 & 0.073184 & 2.5 & 0.073255 \\
\bottomrule
\end{tabular}
\end{table}

\subsection{Identification of the damping}
\label{sec:ident}
\begin{figure}[t]
\centering
\includegraphics[width=\textwidth]{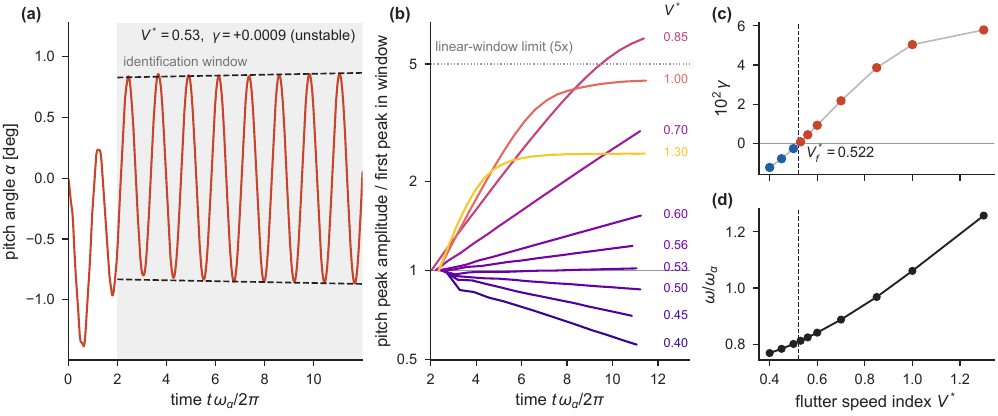}
\caption{Damping identification along a scan of $\Vs$ at $M_\infty=0.85$, $a=-2$. (a) Pitch
history with the identification window (grey) and the fitted envelope (dashed). (b)
Cycle-peak pitch amplitude, normalised by the first peak, for every $\Vs$. (c) Damping
coefficient of the critical mode and the interpolated $\Vf$. (d) Its frequency.}
\label{fig:ident}
\end{figure}

The damping is identified from the pitch and plunge histories by a total-least-squares
matrix pencil \cite{hua1990,jacobson2019}, both channels fitted with common poles, under
three safeguards. (i) \emph{Linear window:} the first two pitch periods are discarded, and
the window ends once the amplitude exceeds five times its initial value or $3^\circ$; a
response that leaves the window is unstable, and its damping is taken from the record
before it left, never from a saturated limit cycle. (ii) \emph{Mode selection:} among the
oscillatory poles between $0.1$ and $5\,\omega_\alpha$ whose energy in the window, summed
over the two channels each scaled by its root-mean-square value, is at least 5\% of that of
the strongest pole, the one with the largest $\sigma$ is the critical mode and gives
$\gamma$. (iii) \emph{Gates:} $R^2\ge0.98$ on both
channels, at least three cycles, and separate fits of the two halves of the window that
each agree in sign with the whole, or lie within 25\% of it; otherwise the run is labelled
indeterminate and not used. Away from neutral stability the labels do not depend on
these settings: varying the energy threshold from 1 to 50\%, the discarded start from one to
four periods, the pencil order from 6 to 12, the channels and the record length changes
$\gamma$ of the three runs that bracket the bottom of the dip by at most
$2.1\times10^{-4}$, and no label (\cref{app:ident}). Only runs within about $0.003$ of
neutral, where two poles of nearly equal frequency can share the window, change label; they
occur near the closure of the fold and are discussed in \cref{sec:family}. \Cref{fig:ident} shows a scan at $M_\infty=0.85$, $a=-2$:
the cycle peaks follow straight lines in logarithmic scale up to $\Vs=0.6$, saturated limit
cycles of about $12^\circ$ at $\Vs=1.0$ and 1.3 never enter the fit, and $\gamma$ crosses
zero once, at $\Vf=0.5224$, where the critical mode has the frequency $0.81\,\omega_\alpha$.

\subsection{Numerical uncertainty and three silent failure modes}
\label{sec:failures}
\Cref{tab:sens} collects the effect of every setting on the flutter speed index at the
bottom of the dip ($M_\infty=0.85$, $a=-2$, $\Vf=0.5224$ on L0). Time step, inner
convergence and record length together stay below 0.2\%. SU2's residual is the flux
balance of a control volume and scales with the cell area, so the residual stop is scaled
in the mesh study ($-6$, $-6.6$, $-7.2$ on L0--L2). The mesh sequence, 0.5224, 0.5183,
0.532, is not monotone, as \citet{mendez2026} also found on five meshes, so no
extrapolation is made and its spread, $-0.8\%$ to $+1.8\%$, is the discretisation
uncertainty.

\begin{table}[t]
\centering
\caption{Sensitivity of the flutter speed index at the bottom of the dip ($M_\infty=0.85$,
$a=-2$; production: mesh L0, 128 steps per pitch period, 12 periods, zero mean incidence,
$\Vf=0.5224$), from runs at the two flutter speed indices that bracket the boundary of each
variant. Time-step rows are relative to 256 steps per period, the others to production.}
\label{tab:sens}
\small
\begin{tabular}{llr}
\toprule
Setting & Variant & Change of $\Vf$ \\
\midrule
Steps per pitch period & 128 (production) & $+0.01\%$ \\
(relative to 256) & 64 & $+0.19\%$ \\
 & 36 (SU2 test case) & $+1.2\%$ \\
 & 32 & label reversed \\
Inner loop & 80 iterations, CFL 10, coupling every 3 & $0.02\%$ \\
 & residual stop $10^{-8}$ & $|\Delta\gamma|\le2.5\times10^{-5}$ \\
Record length & 8 instead of 12 periods & $0.06\%$ \\
Mesh & L1 & $-0.8\%$ \\
 & L2 & $\approx+1.8\%$ \\
Mean incidence & $0.25^\circ$ & $+0.7$ to $+1.2\%$ \\
 & $0.48^\circ$ (impulsive $1^\circ$ start) & $+2.8\%$ \\
 & $0.72^\circ$ & $+4.0\%$ \\
 & $0.94^\circ$ & $+7.4\%$ \\
Amplitude & $\times8.4$ at $0.46^\circ$ mean incidence & about $-0.5\%$ \\
\bottomrule
\end{tabular}
\end{table}

\begin{figure}[t]
\centering
\includegraphics[width=\textwidth]{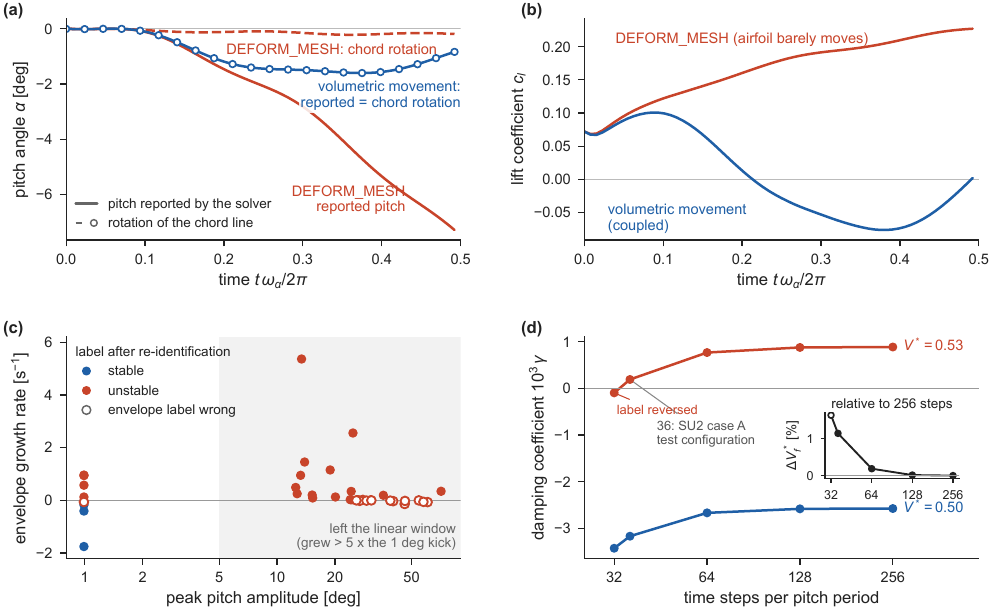}
\caption{Three silent failure modes of damping labels. (a) Reported pitch and geometric
rotation of the chord line, $M_\infty=0.85$, $a=-2$, $\Vs=1.7$, for the two mesh-deformation
paths (they coincide for the volumetric movement); (b) lift coefficient. (c) Growth rate of
an early--late envelope estimator against peak pitch for 40 runs of an earlier campaign;
colours: label after re-identification, open symbols: wrong labels. (d) Damping coefficient
at the two flutter speed indices that bracket the boundary against the number of steps per
pitch period; inset: $\Vf$ relative to the 256-step value (open symbol: extrapolated).}
\label{fig:failures}
\end{figure}

None of the following produces a solver error, a residual divergence or a warning
(\cref{fig:failures}).
\emph{Decoupled mesh motion.} With SU2's typical-section model and its newer
mesh-deformation path (\texttt{DEFORM\_MESH=YES}), the reported pitch evolves but the
aerofoil in the flow barely moves: at $M_\infty=0.85$, $\Vs=1.7$ the pitch reached
$-7.28^\circ$ after half a period while the chord line rotated by at most $0.22^\circ$
(\cref{fig:failures}a,b). The loads stay near their rigid values, every run looks stable and
the learned boundary has no dip. Comparing the surface rotation with the reported pitch
over a few steps detects it; the legacy volumetric movement used here, SU2's own setting
for case~A, matches to $0.001^\circ$. \Cref{app:su2} gives both configurations and a
script that runs this comparison.
\emph{Amplitude saturation.} A growth rate estimated on a limit-cycle plateau is near zero
with an arbitrary sign. In an earlier campaign of 40 runs at $M_\infty=0.85$ evaluated with
an early--late envelope estimator, 31 runs had grown to $12^\circ$--$71^\circ$;
re-identification with the linear window changed 12 labels, all from stable to unstable
(\cref{fig:failures}c), and a surrogate fitted to the old labels shows a spurious
restabilisation.
\emph{Temporal under-resolution.} Dual time stepping converges each step whatever its size,
but a coarse step shifts the phase of the aerodynamic response, which near the boundary
sets the sign of the damping. With 32 steps per pitch period the run at $\Vs=0.53$, 1.5\%
above the boundary, is labelled stable with both halves agreeing (\cref{fig:failures}d);
\citet{mendez2026} found such a reversal with 50 steps. The coarse step always errs in the
non-conservative direction.

\subsection{Mean incidence and excitation amplitude}
\label{sec:incidence}
\begin{figure}[t]
\centering
\includegraphics[width=\textwidth]{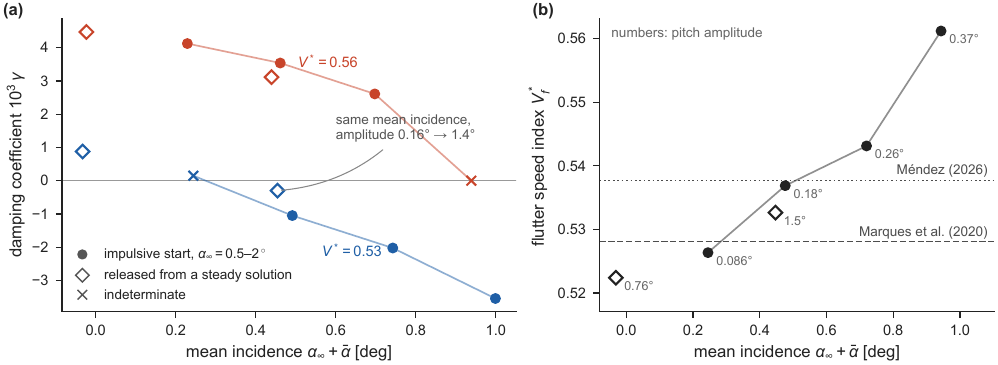}
\caption{Effect of the start of the motion at $M_\infty=0.85$, $a=-2$. (a) Damping coefficient
at $\Vs=0.53$ and 0.56 against the mean incidence (freestream incidence plus mean pitch);
filled: impulsive starts, open diamonds: release from a steady solution, crosses:
indeterminate; the annotated pair differs eightfold in amplitude. (b) Interpolated $\Vf$
against the mean incidence (numbers: pitch amplitude), with the published values of
\citet{marques2020} and \citet{mendez2026} ($1^\circ$ incidence).}
\label{fig:incidence}
\end{figure}

Three angles must be distinguished. The \emph{freestream incidence} $\alpha_\infty$ is
the flow angle of the unsteady run. The \emph{initial perturbation} is the departure of the
initial flow and structure from equilibrium at that incidence; in all our runs the section
starts undeflected and at rest, so the perturbation is aerodynamic, and no initial pitch is
prescribed. The \emph{mean incidence} $\alpha_\infty+\bar\alpha$, with $\bar\alpha$ the mean
pitch, is the incidence about which the section oscillates once its static deflection has
developed. Starting a run from a freestream incidence mixes two effects: a static deflection,
so that the section oscillates about a non-zero mean incidence, and the oscillation
amplitude. We separated them at $M_\infty=0.85$, $a=-2$ with impulsive starts from
uniform flow at $\alpha_\infty=$\,$0.25^\circ$--$2^\circ$,
release from the steady $1^\circ$ solution into $\alpha_\infty=0$,
which perturbs the flow but leaves zero mean incidence, and release from the steady
$2^\circ$ solution into $\alpha_\infty=$\,$1^\circ$, which keeps the mean incidence of
the impulsive $1^\circ$ start at eight times its amplitude (\cref{fig:incidence}). The
nose-down twist about the forward elastic axis roughly halves the incidence: an
impulsive start at $\alpha_\infty=1^\circ$ oscillates about $0.48^\circ$. $\Vf$ rises monotonically with the mean
incidence (\cref{tab:sens}), while an 8.4-fold amplitude changes it by about $-0.5\%$.
The direction agrees with unsteady Reynolds-averaged results at larger preset angles
\cite{gao2018} and with small-disturbance computations \cite{edwards1983}; new is the size
at the incidences used to excite benchmarks: $0.25^\circ$ moves $\Vf$ several times more
than the time step or the inner convergence, and $0.5^\circ$ more than any numerical
setting, including the mesh. 
\Cref{tab:published} compares the two published values at this condition with ours.
Neither study reports the mean incidence of its unsteady run, so the comparison cannot
establish equivalent conditions. \citet{mendez2026} states an angle of attack of $1^\circ$.
Read as the freestream incidence, it corresponds to our runs at $\alpha_\infty=1^\circ$,
whose $\Vf=0.5369$ is within 0.15\% of their 0.5377; since the amplitude matters little, the
start then matters little too. Read as the mean incidence including the static deflection,
it would correspond to a flutter speed index about 7\% above our zero-incidence value
(\cref{tab:sens}), not the 2.9\% observed. \citet{marques2020}
start from the steady solution at $1^\circ$ and do not state the incidence of the unsteady
run; their 0.5281 lies 1.1\% above our zero-mean-incidence value, inside our mesh spread, and
their mesh, time step and record differ from ours. The 1.8\% between the two published values
therefore has the size and the sign of a difference in incidence protocol, and we offer this
as a physically supported interpretation, not as a demonstration that the protocols differed
in this way. All results below use zero mean incidence.

\begin{table}[t]
\centering
\caption{Published flutter speed indices at the bottom of the dip ($M_\infty=0.85$,
$a=-2$, SU2 Euler simulations) and the conditions under which they were obtained, as
stated in each study; ``n.s.'': not stated.}
\label{tab:published}
\small
\begin{tabular}{@{}>{\rr}p{25mm}>{\rr}p{33mm}>{\rr}p{33mm}>{\rr}p{52mm}@{}}
\toprule
 & \citet{marques2020} & \citet{mendez2026} & This work \\
\midrule
Mesh points & 8\,606 & 4\,728--21\,294 (five meshes) & 5\,233 (L0), 20\,682 (L1), 82\,228 (L2) \\
Steps per pitch period & 100 & 100 & 128 \\
Record & 10 periods, first 30\% discarded & 60 periods & 12 periods, first 2 discarded \\
Identification & matrix pencil, pitch and plunge separately & matrix pencil, pitch, plunge,
$c_l$ and $c_m$ & matrix pencil, common poles of pitch and plunge \\
Initial flow & steady solution at $1^\circ$ & n.s. & (a) steady solution at $1^\circ$;
(b) uniform flow \\
Freestream incidence & n.s. & $1^\circ$ (stated as angle of attack) & (a) $0$; (b) $1^\circ$ \\
Mean incidence & n.s. & n.s. & (a) $0$; (b) $0.48^\circ$ \\
$\Vf$ & 0.5281 & 0.5377 (finest mesh) & (a) 0.5224 (L0), 0.5183 (L1), $\approx0.532$ (L2);
(b) 0.5369 (L0) \\
\bottomrule
\end{tabular}
\end{table}

\section{Learning the boundary with calibrated uncertainty}
\label{sec:learning}
The boundary is learned in three steps: a Gaussian-process surrogate of the damping
coefficient (\cref{sec:gp}), batches of simulations placed where its zero contour is most
uncertain (\cref{sec:al}), and a calibration of its uncertainty on independent simulations
(\cref{sec:conformal}). The implementation is part of the code archive (Data availability):
module \texttt{flutterrom.\allowbreak active\_\allowbreak learning.\allowbreak contour} and script
\texttt{su2\_\allowbreak contour\_\allowbreak campaign.py}, with scikit-learn~1.9 for the Gaussian process.

\subsection{Surrogate model}
\label{sec:gp}
Following \citet{marques2020}, a Gaussian process \cite{rasmussen2006} is fitted to
$y=\tanh(\gamma/s)$, $s=0.03$; the transform keeps the zero set and saturates the
large damping far from the boundary, which would otherwise dominate the fit. Its covariance
is $k(\mathbf{x},\mathbf{x}')=\sigma_f^2\,k_{5/2}(r)+\sigma_n^2\,\delta_{\mathbf{x}\mathbf{x}'}$:
an anisotropic Mat\'ern-5/2 term, with $r$ the distance between the inputs scaled by one
length scale per input, and a white-noise term that absorbs the scatter of the identified
damping. The inputs are scaled to the unit cube by the fixed bounds of the domain, not by
the data, so that the scaling does not drift as samples cluster on the boundary, and the
targets are standardised. The hyperparameters maximise the log marginal likelihood within
$\sigma_f^2\in[10^{-3},10^{3}]$, length scales in $[10^{-2},10]$ and
$\sigma_n^2\in[10^{-8},10^{-1}]$ (standardised units), from an initial length scale of 0.3
and four random restarts during a campaign (eight for the final maps). The boundary
estimate is the zero contour of the posterior mean $m(\mathbf{x})$, folds included,
extracted on a regular grid ($221\times331$ points over $(M_\infty,\Vs)$ for case~A); $s_y$
denotes the posterior standard deviation.

Runs labelled indeterminate (\cref{sec:ident}) are stored but not fitted. A run
that leaves the linear window within a few cycles is certainly unstable, but its
short-window damping is unreliable; its target is therefore bounded below by
$\tanh(\gamma_{\min}/s)$, $\gamma_{\min}=\ln2/(2\pi n)$, the damping of a mode that at least
doubles in the $n$ observed cycles, and a run labelled unstable never enters with
$y<10^{-3}$. Without this bound such runs enter as nearly neutral and create closed
spurious contours inside the unstable region, which five additional simulations confirmed
to be unstable.

\subsection{Batch active learning}
\label{sec:al}
The candidates are a regular grid of the domain: 51 Mach numbers by 96 flutter speed
indices for case~A (4\,896 points), and $51\times21\times96$ points over Mach number,
elastic axis and $\Vs$ for the family. The initial design is a Latin hypercube, of 16 runs
for case~A and 24 for the family. Each round refits the surrogate to every labelled
run and selects one condition per processor core, eight in all, which then run
concurrently as independent serial SU2 processes. A candidate is scored by the entropy
$H=-\sum p\ln p$ of its posterior probabilities $p$ of being stable
($y<-\epsilon$), near-neutral ($|y|\le\epsilon$) or
unstable ($y>\epsilon$), with $\epsilon=2s_y$, the local term of the contour
entropy of \citet{marques2020}. It is largest near the predicted contour where the
posterior is uncertain, and small both far from the contour and where the contour is
already pinned by data. The batch is built greedily by updating
the posterior variance as if each chosen point had been observed at its mean
(kriging believer; the mean is unchanged, so a rank-one update of the variance suffices)
and by excluding candidates closer than 0.02 in the unit cube to a chosen point. Case~A
used eight rounds (80 runs in all) and the family fourteen (136 runs), numbers fixed in
advance by the computing time available; the mean contour entropy over the domain was
recorded in every round. The random baseline of \cref{sec:efficiency} replaces the
selection by draws from the same candidate grid, without replacement within a round, so
that a condition can recur in a later round.

\subsection{Conformal calibration and stability calls}
\label{sec:conformal}
Split conformal prediction \cite{lei2018} turns the posterior into calibrated calls.
For a calibration set of $n$ runs not used for fitting, the scores
$r_i=|y_i-m(\mathbf{x}_i)|/s_y(\mathbf{x}_i)$ are computed with the fitted surrogate, and $q$
is the $\lceil(n+1)(1-\alpha)\rceil$-th smallest of them. A condition is called stable
where $m+q\,s_y<0$ and unstable where $m-q\,s_y>0$; inside the band $|m|\le q\,s_y$ no call
is made. Since active learning samples near the boundary, the calibration set is drawn
uniformly over the domain.

If a new run $(\mathbf{x},y)$ is exchangeable with the calibration runs, in particular
if $\mathbf{x}$ is drawn from the same uniform distribution, then
$P\{|y-m(\mathbf{x})|\le q\,s_y(\mathbf{x})\}\ge1-\alpha$, the probability being taken over
the calibration set and the new run together \cite{lei2018}. A wrong call needs $y$ outside
the band, so the probability that a run receives a call and the call is wrong is at most
$\alpha$. This statement has three limits. It is marginal: it holds on average over
conditions distributed like the calibration set, not at each condition, and not along the
boundary, where the conditions that matter are not distributed like that. It bounds the
joint probability of a call and an error; the share of wrong calls among the calls made,
which we report, is bounded only by $\alpha$ divided by the fraction of runs that receive a
call. And it attaches no probability to a statement that combines several calls, such as
the existence of a fold, which needs a confidently unstable condition below a confidently
stable one on the same line of constant Mach number and elastic axis; such a statement is
supported by calls that each have the rate above, but is not guaranteed jointly. The
empirical label accuracy, the share of independent runs for which the sign of $m$ is right,
is a third quantity: it uses no band and carries no guarantee, and we report it as a test
of the map.

\Cref{tab:datasets} lists the simulations and their roles. For case~A the random
campaign provides the independent runs: its 64 distinct runs beyond the shared initial
design are never fitted by the active-learning surrogate. All 64 give the band shown in
the figures and test the labels; the coverage and call rates are measured on 200 random
halves, calibrating on one half and testing on the other, so that every reported coverage
and call rate comes from runs held out from both fitting and calibration. The 40 uniform
runs of the family play the same role there.

\begin{table}[t]
\centering
\caption{Simulation sets (mesh L0, zero mean incidence) and their roles. Case~A
domain: $0.70\le M_\infty\le0.92$, $0.35\le\Vs\le2$; family domain: $0.80\le M_\infty\le0.92$,
$-2.5\le a\le-1$, $0.25\le\Vs\le2$.}
\label{tab:datasets}
\small
\begin{tabular}{@{}>{\rr}p{30mm}>{\rr}p{55mm}>{\rr}p{65mm}@{}}
\toprule
Set & Runs & Role \\
\midrule
Case~A, active learning & 80: Latin hypercube of 16, eight batches of 8 & fits the case~A
surrogate (\cref{fig:boundaryA}) \\
Case~A, random & 81: the same 16, then 65 random draws from the candidate grid, one of them
repeated; 64 distinct beyond the shared design & independent set of the active-learning
surrogate: label test, calibration ($q$), held-out coverage and calls (200 halves); fits the
random-sampling surrogates (\cref{fig:efficiency}b, 40 splits with 24 held out) \\
Case~A, union & 144 distinct & reference map of \cref{fig:efficiency}a \\
Family, active learning & 136: Latin hypercube of 24, fourteen batches of 8 & fits the
family surrogate \\
Family, reused and checks & 47 case~A active-learning runs inside the family domain, and
ten checks & fit the family surrogate (193 runs in all) \\
Family, calibration & 40, uniform over the domain & calibration ($q$), held-out coverage and
calls (200 halves); never fitted \\
\bottomrule
\end{tabular}
\end{table}

\section{Results}
\label{sec:results}

\subsection{The case~A boundary}
\label{sec:caseA}
\begin{figure}[t]
\centering
\includegraphics[width=\textwidth]{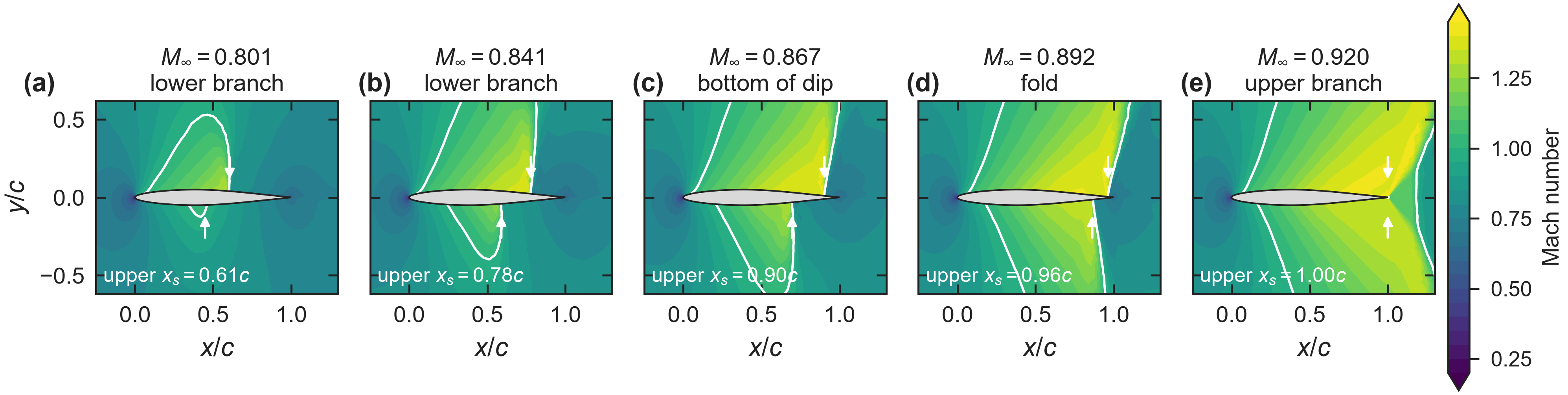}
\caption{Flow along the case~A boundary: steady Mach number at $1^\circ$ incidence, the
state from which every simulation is released (mesh L0), at five Mach numbers on the lower
branch, at the bottom of the dip, in the fold and on the restabilisation branch. White:
sonic line; arrows: shock positions on the two surfaces; $x_s$: upper-surface shock.}
\label{fig:regimes}
\end{figure}

\begin{figure}[t]
\centering
\includegraphics[width=\textwidth]{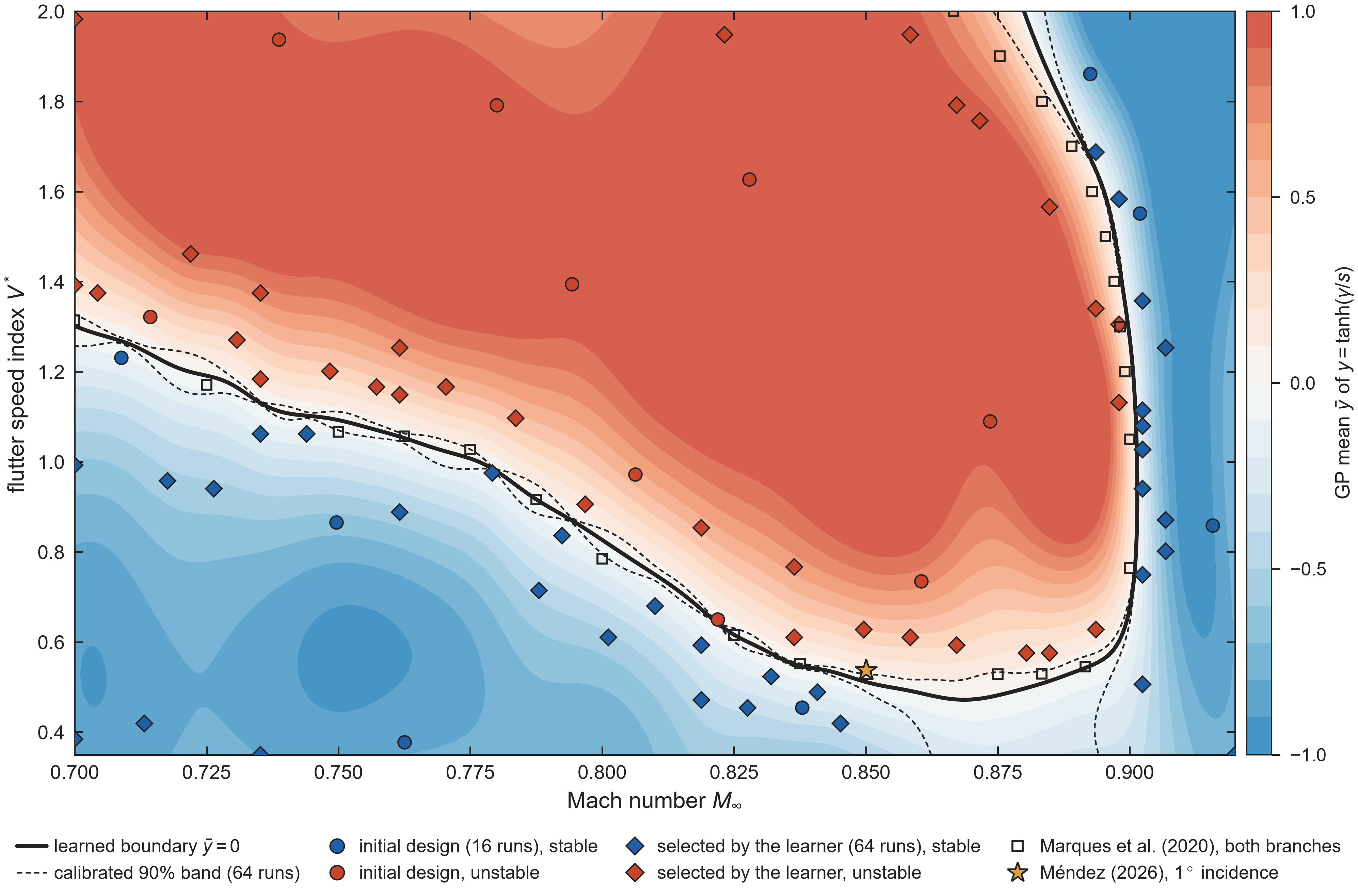}
\caption{The case~A flutter boundary learned from 80 coupled Euler simulations (mesh L0,
zero mean incidence). Colour: posterior mean $\bar y$ of $y=\tanh(\gamma/s)$; thick line:
learned boundary $\bar y=0$; dashed: calibrated 90\% band from 64
independent random simulations ($q=1.59$). Circles: initial design; diamonds: chosen by the learner; open
squares: \citet{marques2020}; star: \citet{mendez2026} at $1^\circ$ incidence.}
\label{fig:boundaryA}
\end{figure}

\Cref{fig:regimes} shows the flow along the Mach range of the campaign: the supersonic
region grows and its shock moves aft, from $0.61c$ at $M_\infty=0.80$ to the trailing edge
at 0.92. The learned boundary (\cref{fig:boundaryA}), from a Latin-hypercube design of 16
simulations and eight batches of eight (80 in all, none failed or indeterminate), follows
this sequence. The lower branch descends from $\Vf=1.30$ at $M_\infty=0.70$ to the bottom
of the dip, $\Vf=0.472$ at $M_\infty=0.868$, and turns up steeply towards
$M_\infty\approx0.90$, where it meets the nearly vertical restabilisation branch: up to
$\Vs=1.8$ the section is unstable on its low-Mach side and stable beyond
$M_\infty\approx0.89$--0.90. Of the 64 simulations placed by the learner, 25 lie within
$\pm0.005$ in Mach number and $\pm0.05$ in $\Vs$ of the final boundary, against 3 of the
16 initial ones.

Against the bisection reference of \citet{marques2020}, the lower branch agrees to within
2.4\% for $M_\infty=0.70$--0.84 (5.4\% at 0.80, where their boundary falls steeply), and
the restabilisation branch to within 0.0024 in Mach number up to $\Vs=1.7$. At the bottom
of the dip, $M_\infty=0.85$--0.88, the learned boundary lies 3--9\% below the reference,
where the learner placed no stable run near the boundary; at $M_\infty=0.85$ it gives
0.511 against the bisection value 0.5224. The calibrated band (\cref{sec:efficiency})
spans $0.489\le\Vs\le0.528$ there, containing both values, and reaches the bottom of the
domain between $M_\infty=0.868$ and 0.892: the band flags the floor of the dip as the part
that needs more simulations. Eleven of the 14 reference points on the lower branch lie
inside it. On the restabilisation branch the band is only 0.0002--0.0005 wide in Mach number
and the reference lies 0.0012--0.0024 lower; the band measures how well the surrogate
represents our simulations, not differences of mesh, solver or start protocol between
studies, which amount to about one per cent at the dip (\cref{tab:sens}).

\subsection{Learning efficiency and calibrated stability calls}
\label{sec:efficiency}
\begin{figure}[t]
\centering
\includegraphics[width=\textwidth]{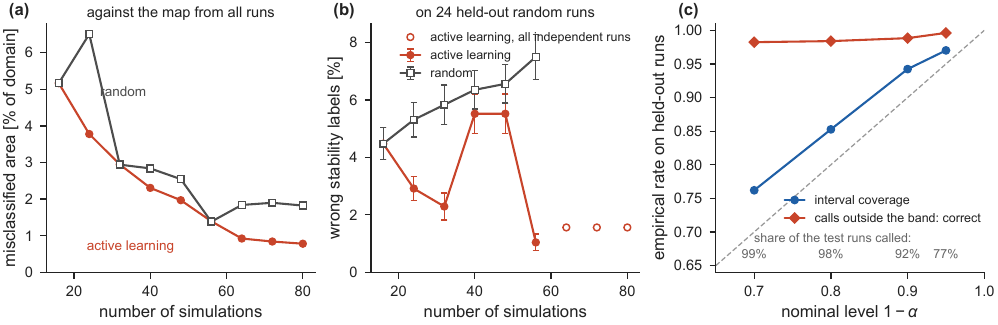}
\caption{Learning efficiency and calibration for case~A. (a) Fraction of the domain where
the map after $n$ runs disagrees with the map from all runs, for active learning and random
sampling from the same 16-run design. (b) Wrongly labelled held-out random runs (mean and
standard error, 40 splits); open circles: active learning scored on all
64. (c) Empirical
coverage and correct calls outside the band on held-out halves (200 splits) against the
nominal level; numbers: share of runs that receive a call.}
\label{fig:efficiency}
\end{figure}

A second campaign of the same cost started from the same 16 runs and added simulations
drawn at random from the candidate grid, 81 in all;
one condition was drawn twice, so 64
distinct runs lie beyond the shared design. They are never fitted by the active-learning
surrogate and form an unbiased test set for it and
the calibration set of its band (\cref{tab:datasets}). Against the map fitted to all
144 distinct runs (16 shared, 64 active-learning and 64 random;
\cref{fig:efficiency}a) the disagreement falls to 0.8\%
of the domain after 80 active-learning runs, while random sampling stays near 1.8\%. On
held-out runs only (\cref{fig:efficiency}b; 40 splits, 24 of the 64
held out), random sampling leaves 5--8\% wrongly labelled from 24 to 56
runs, and active learning reaches 1.0\% at 56 runs and then labels
63 of the 64 independent runs correctly with 64, 72 and 80
runs. The one it gets
wrong is stable, at $M_\infty=0.872$, $\Vs=1.98$, at the top of the restabilisation branch,
and lies outside the band ($|m|/s_y=2.5$): one wrong call, the kind of error whose rate the
calibration bounds. Split 200 times into calibration and test halves
(\cref{fig:efficiency}c), the coverage of $m\pm q\,s_y$ is 0.970, 0.942, 0.853 and 0.762 at nominal 0.95, 0.90, 0.80 and 0.70; 77, 92, 98 and 99\% of the test runs receive a
call, and 0.4, 1.2, 1.6 and 1.8\% of the calls made are wrong. The joint rate of a call and an
error, the quantity bounded by $\alpha$ (\cref{sec:conformal}), is smaller still.

\subsection{Energy transfer at the bottom of the dip}
\label{sec:energy}
\begin{figure}[t]
\centering
\includegraphics[width=\textwidth]{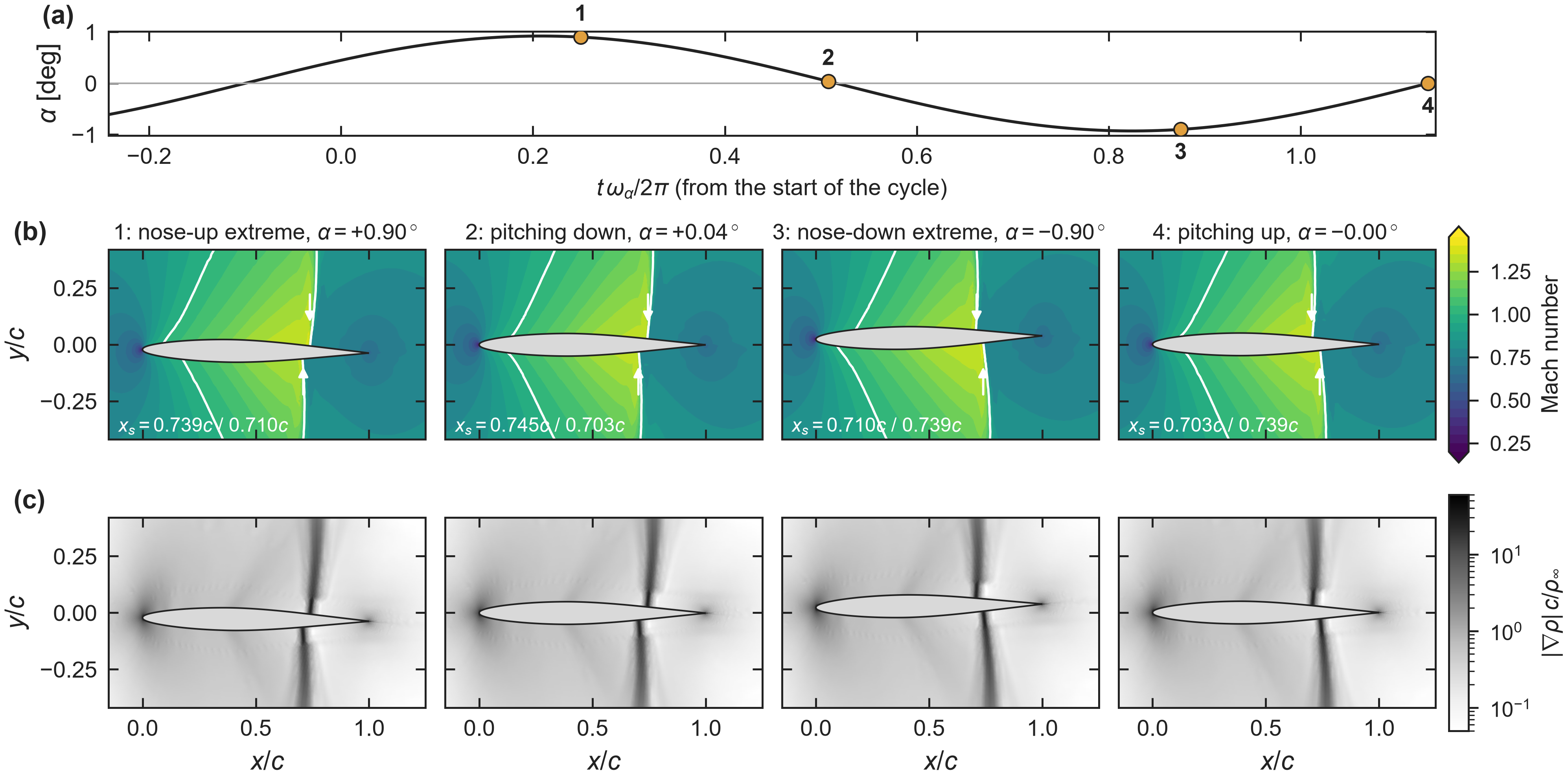}
\caption{One flutter cycle at the bottom of the dip: $M_\infty=0.85$, $a=-2$, $\Vs=0.53$
(unstable), mesh L1. (a) Pitch over the last cycle and the four instants shown. (b) Mach
number with the sonic line (white), the shocks (arrows) and their positions on the upper
and lower surfaces. (c) Numerical schlieren $|\nabla\rho|\,c/\rho_\infty$. The shock lags the
pitch: of the four instants, the upper shock is furthest aft at 2, after the nose-up
extreme.}
\label{fig:cycle}
\end{figure}

Two simulations on mesh L1 at $M_\infty=0.85$, $a=-2$ bracket the boundary ($\Vf=0.518$ on
this mesh): $\Vs=0.50$ is stable ($\gamma=-0.0022$) and
$0.53$ unstable ($+0.0015$). Volume and surface solutions were written every 16 and every 2 steps.
\Cref{fig:cycle} shows one flutter cycle of the unstable run. Each surface carries a
supersonic region closed by a nearly normal shock near $0.72c$, which moves by about
$\pm0.02c$ for $0.9^\circ$ of pitch, in antiphase on the two surfaces, and reaches its
aftmost position only after the nose-up extreme.

\begin{figure}[t]
\centering
\includegraphics[width=\textwidth]{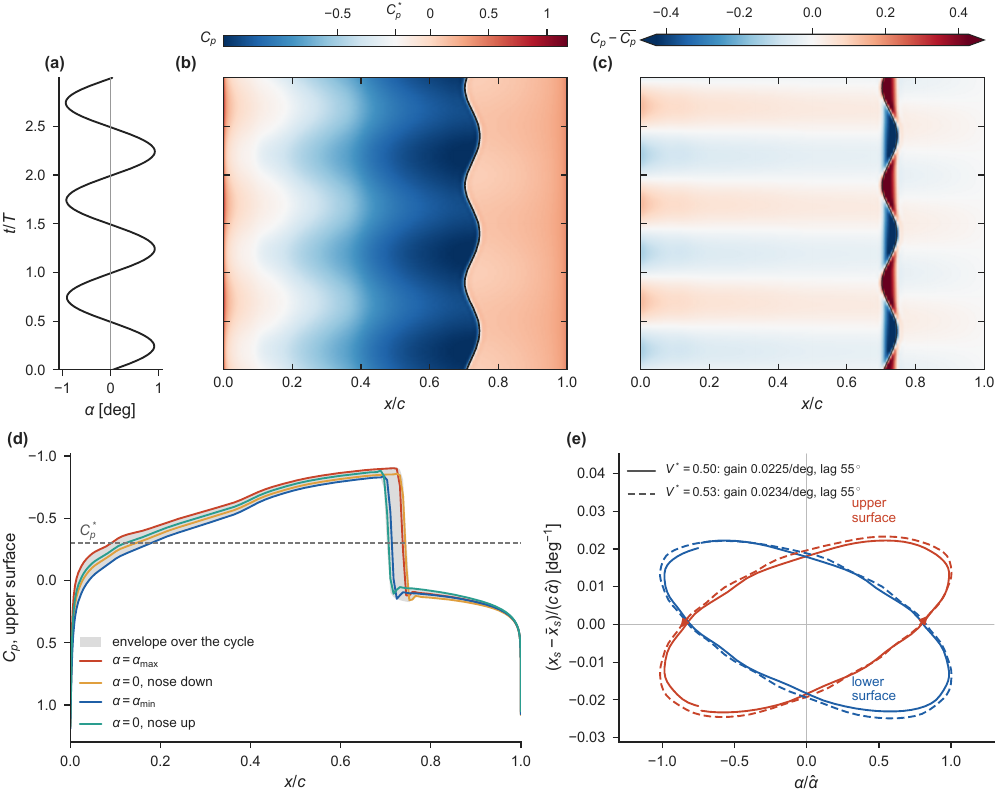}
\caption{Surface pressure and shock motion at $M_\infty=0.85$, $a=-2$ (mesh L1), last three
cycles. (a) Pitch; (b) space--time map of the upper-surface $C_p$ at $\Vs=0.53$, centred on
$C_p^*$, with the shock track; (c) its unsteady part; (d) $C_p$ at four phases and its
envelope; (e) shock position against pitch, normalised by the pitch amplitude, below and
above the boundary. The weak inflections near $x/c=0.1$ and 0.4 follow curvature changes
of the tabulated section.}
\label{fig:surface}
\end{figure}

Tracked as the point where the surface pressure rises through $C_p^*$ (\cref{fig:surface}),
the shock has the mean position $0.724c$ in both runs, moves aft by $0.0225c$ and $0.0234c$
per degree of nose-up pitch, and lags the pitch by $55^\circ$ in both. Its loops normalised
by the pitch amplitude coincide (\cref{fig:surface}e), and the mode is the same:
$|h_1|/(b|\alpha_1|)=1.844$ with the plunge within $1.5^\circ$ of the pitch phase, so the
section pitches about a point $1.42c$ ahead of the leading edge, as in Isogai's
small-disturbance solution at $M_\infty=0.825$ (ratio 1.86, zero phase \cite{isogai1981}).
The mean shock lies in the band 0.6--0.8 chord that \citet{bendiksen2011} identifies as the
most dangerous. What changes across the boundary is the reduced frequency
$k=\omega b/U_\infty$, from 0.207 to 0.198. The supersonic region responds almost in phase
with the motion, as in quasi-steady flow, and the shock stays on the surface and moves nearly
sinusoidally throughout the cycle, which is type~A shock motion in the classification of
\citet{tijdeman1980}.

\begin{figure}[t]
\centering
\includegraphics[width=\textwidth]{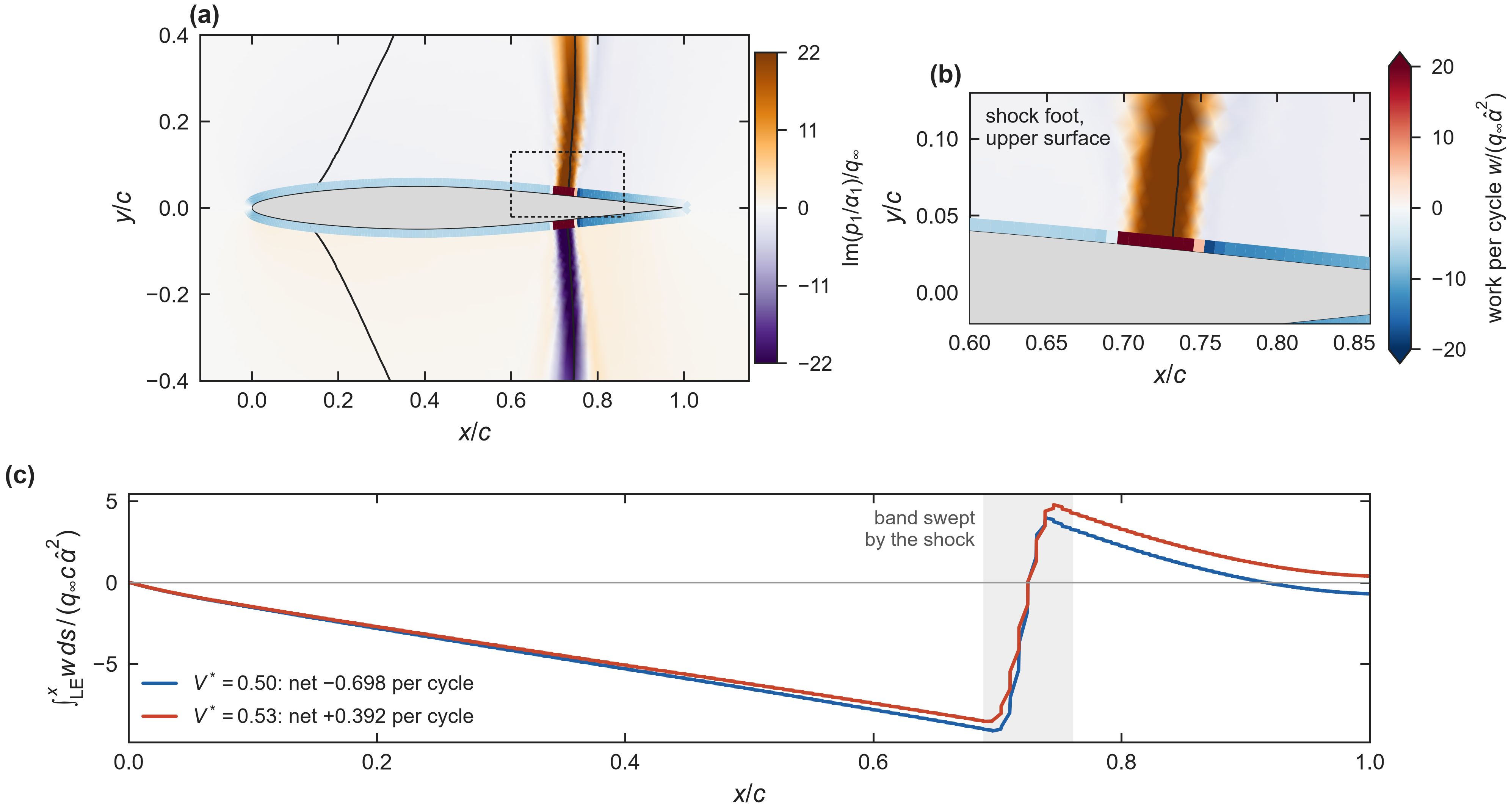}
\caption{Where the flow feeds the flutter mode, $M_\infty=0.85$, $a=-2$, mesh L1, last four
cycles. (a) Quadrature part of the pressure at the critical pole,
${\rm Im}(p_1/\alpha_1)/q_\infty$, at $\Vs=0.53$, with the mean sonic line (black); the
aerofoil contour is coloured by the local work per cycle $w/(q_\infty\hat\alpha^2)$ done by
the flow (red: into the structure). (b) The shock foot on the upper surface. (c) Running
integral of $w$ from the leading edge over both surfaces, below and above the boundary,
with the net work per cycle.}
\label{fig:energy}
\end{figure}

The local work of the unsteady pressure per cycle,
$w=-\oint p'\,\mathbf n\cdot\mathbf v_{\mathrm{wall}}\,\mathrm dt$, is evaluated from the
components of $p$ and $\mathbf n\cdot\mathbf v_{\mathrm{wall}}$ at the critical pole, fitted at
every surface point with all the other poles of the motion; a single-frequency fit or a
plain window integral picks up the slow release transient and changes the net by up to
20\%, whereas every term below changes by less than 0.04 between windows of three and five
cycles. \Cref{fig:energy}a shows where the energy enters: the pressure component in
quadrature with the pitch, the part that does work, is concentrated in the band swept by
the shock, and the surface there is red. Three independent evaluations of the net work per
cycle agree within 3.5\% (normalised by $q_\infty c\,\hat\alpha^2$): the surface integral
gives $-0.698$ and $+0.392$ at $\Vs=0.50$ and 0.53, the solver's lift and moment $-0.708$
and $+0.391$, and the growth rate, through $W=4\pi\gamma\langle E\rangle$
with the cycle-mean structural energy $\langle E\rangle$, $-0.687$ and $+0.405$.

Per cycle, the surface ahead of the shock removes 9.02 at $\Vs=0.50$ and
8.54 at 0.53, the
surface behind it 3.96 and 3.89, and the band swept by the shock ($0.69\le x/c\le0.76$)
supplies 12.29 and 12.82 (\cref{fig:energy}c). The net is 3--6\% of the energy passing
through the shock band. The band contains the shock excursion, about $0.70$--$0.75c$,
and the few cells over which the captured shock is spread, but also part of the smooth
pressure field on which the shock rides. To separate the two, the work density of the
smooth field is interpolated linearly across the band, on each surface, between its mean
values in strips $0.03c$ wide on either side. The band integral of this interpolant, $-1.42$
and $-1.38$, is the smooth part, and the remainder, 13.71 and 14.20, is the work of the
shock. The shock work changes by at most 1.6\% when the band is widened to $0.68$--$0.77$,
$0.67$--$0.78$ or $0.65$--$0.80$, whereas the band integral varies from 10.7 to 12.8 as the
band takes in more of the damping surface (\cref{tab:shockband}); only a band narrower than
the excursion, $0.70$--$0.75$, cuts into it.
Its change of sign, $+1.09$ per cycle, comes in comparable
parts from a larger shock input ($+0.49$, or 45\% of
the change, 42--45\% for the wider bands; the shock work rises by 3.6\%, with the shock
excursion per degree, 4.0\%) and a smaller loss ahead of the shock
($+0.48$, falling by 5.3\%, close to the 4.4\% fall
of $k$, as for a quasi-steady response); the smooth parts of the band and of the surface
behind it add the remaining 11\%. With the Mach number
fixed, and the mode, the shock phase and the normalised shock loop nearly unchanged, the
reduced frequency is the one aerodynamic parameter that changes across this boundary, and
it acts on both parts at once. The lower branch is therefore crossed through the reduced
frequency, which shifts the balance between the input of the moving shock and the
quasi-steady damping of the rest of the surface. It is the controlling parameter of the
crossing, not its unique cause: the shock input is what brings the net work close enough to
zero for a 4\% change of $k$ to reverse its sign, and neither part alone accounts for the
change. The two surfaces contribute equally to within 0.07 per cycle.

\subsection{Elastic-axis family and the emergence of the fold}
\label{sec:family}
\begin{figure}[t]
\centering
\includegraphics[width=\textwidth]{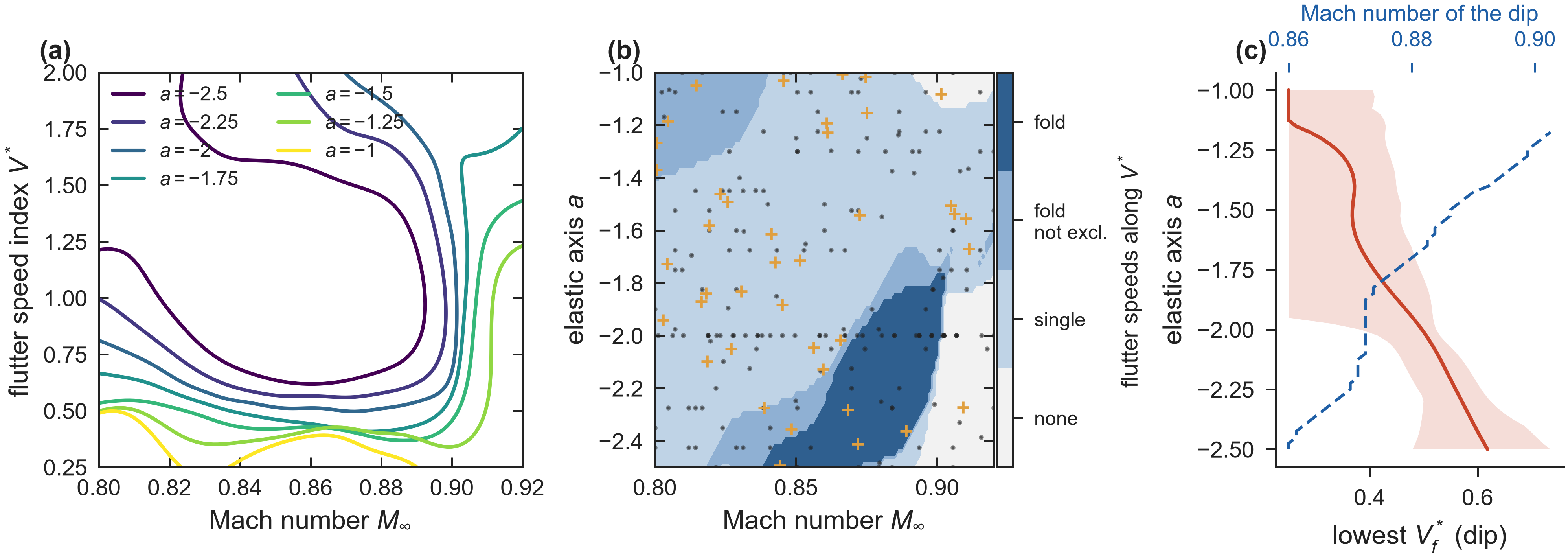}
\caption{The elastic-axis family from 193 coupled simulations over $(M_\infty,a,\Vs)$, mesh
L0, zero mean incidence, calibrated by 40 independent random simulations ($q=1.535$ at 90\%).
(a) Learned boundary for seven elastic-axis positions. (b) Flutter speeds along $\Vs$ at each
$(M_\infty,a)$: none, a single one, a fold not excluded by the calibrated band, and a
calibrated fold (a confidently unstable $\Vs$ below a confidently stable one); dots:
simulations, crosses: calibration set. (c) Bottom of the dip with its calibrated band, and
its Mach number, against the elastic axis.}
\label{fig:family}
\end{figure}

The elastic axis was varied from $a=-2.5$ to $-1.0$ (from three quarters of a chord ahead of
the leading edge to the leading edge) over $M_\infty\in[0.80,0.92]$ and $\Vs\in[0.25,2.0]$,
the other structural parameters at their case~A values: a 24-run design and fourteen
batches of eight, 136 simulations, plus the 47 case~A runs inside this domain and ten
checks (193 runs, \cref{fig:family}). Of five checks placed where the map makes its main
statements, four confirmed the predicted label; the fifth, and four more, lay in small
closed stable contours at high $\Vs$ that were the capped-run artefact of
\cref{sec:learning} and vanished once those runs were bounded from below. Forty further
simulations, drawn uniformly over the domain and not used for fitting, calibrate the band
as for case~A. The sign of the posterior mean is right for all 40, and with all of them
$q=1.535$ at $\alpha=0.1$. Split into halves 200 times, they give coverages of 0.958, 0.958,
0.870 and 0.769 at nominal levels 0.95, 0.90, 0.80 and 0.70, and none of the stability calls
made outside the band, for 70--93\% of the test runs, is wrong.

The bottom of the dip falls steadily as the elastic axis moves aft (\cref{fig:family}c):
$\Vf=0.62$ at $a=-2.5$, 0.46 at $-1.9$, 0.36 at $-1.3$, and the lower edge of the domain,
0.25, at $-1.0$, while its Mach number moves from 0.86 to 0.90. In this inviscid
model the fold, two or more
flutter speeds at one Mach number, exists only for the forward elastic axes
(\cref{fig:family}b): it spans $0.824\le M_\infty\le0.891$ at $a=-2.5$, 0.860--0.899 at
$-2.2$ and 0.878--0.901 at $-1.9$, and closes at $a\approx-1.75$ near $M_\infty=0.90$. As
the axis moves forward, the restabilisation branch stays near $M_\infty=0.89$--0.90 while
the lower branch rises and its dip moves to lower Mach numbers; the two separate and the
range of multiple flutter speeds widens. For aft axes the flutter region is no longer
closed at high $\Vs$, and a single flutter speed remains at each Mach number.

The calibrated band turns these statements into stability calls with a bounded
marginal error rate
(\cref{fig:family}b). A fold is supported by calibrated calls, with a
confidently unstable flutter speed index
below a confidently stable one, for $a\le-1.78$; for $-1.75\le a\le-1.70$ the band cannot
decide, and for $a\ge-1.65$ a fold is excluded near $M_\infty=0.90$. The closure therefore
lies between $a=-1.78$ and $-1.65$ in this model. Each of these statements rests on two
calls at specific conditions, each with the marginal error rate of \cref{sec:conformal}; no
joint probability is attached to the existence of the fold, and the interval is checked by
simulations below. Elsewhere a fold is not excluded only at the edge of the
domain, $M_\infty\ge0.918$, and in the corner $M_\infty<0.84$, $a>-1.4$, where few simulations
lie at high $\Vs$. The band of the dip is wide (0.43--0.54 at $a=-2.05$, and reaching the
lower edge of the domain for $a\ge-1.9$; \cref{fig:family}c): the three-parameter map fixes
the trend of the dip, not its value. At $a=-2$ it puts the dip at 0.51 against 0.47 from the
denser two-parameter campaign; a single axis is better mapped in two parameters.

The closure was checked on the finer mesh L1 at $M_\infty=0.89$, with the scaled residual
stop of \cref{sec:failures}. At $a=-1.9$ the fold is reproduced: the section is unstable at
$\Vs=1.2$ and stable at $\Vs=2.0$ ($\gamma=-0.046$). At $a=-1.6$ it is unstable at $\Vs=2.0$,
leaving the linear window, so there is no restabilisation. At $a=-1.75$, $\Vs=2.0$ is
neutral to within $\gamma=+0.0002$, too close for a confident label: the two halves of
its window disagree in sign ($+0.0023$ and $-0.0012$), and four-period windows fall from
$+0.0024$ to $+0.0009$ along the record. The fold therefore
closes at $a\approx-1.75$ on both meshes, in inviscid flow.

The runs of mesh L0 on which the closure interval rests were re-identified with the
variants of \cref{sec:ident} (\cref{app:ident}). Those at $a=-1.825$ and $-1.78$, near
$M_\infty=0.90$ and $\Vs=1.7$, have a single dominant pole and are stable for every energy
threshold, discarded start, pencil order, channel choice and record length of 8, 10 or 12
periods. The run at $a=-1.75$ is within 0.003 of neutral: two poles of nearly equal frequency
share its window, the stronger neutral ($\gamma=-0.0001$) and the weaker, with 36\% of its
energy, growing ($+0.0018$), so that it is labelled unstable for energy thresholds up to
20\% and indeterminate at 50\%, and unstable for every record length. Within records of 8 to
12 periods the identification therefore does not move the interval; only the label at
$a=-1.75$, where the band is already undecided, depends on it.

The closure was also verified directly at $M_\infty=0.89$, $\Vs=2.0$, on the
restabilisation branch (\cref{tab:closurecheck}a). With the case~A structure about the axis
the section is stable at $a=-1.80$ and unstable from $-1.70$ aft on both meshes, with both
time steps and for every record length tested, while at $\Vs=1.2$ it flutters at $a=-1.75$
and $-1.6$ ($\gamma=+0.052$ and $+0.058$). Halving the time step changes $\gamma$ by at most
$1.5\times10^{-4}$; mesh L1 changes it by $-0.010$ at $a=-1.80$ and $+0.0014$ at $-1.70$,
without changing a label. Between the two, $a=-1.75$ is within 0.005 of neutral: on L0 it
is unstable with both time steps and every record length, with $\gamma$ falling from
$+0.0043$ over 8 periods to $+0.0013$ over 20, and on L1 it is neutral to within 0.0002.
The sign change at this condition therefore lies near $a=-1.76$ on L0 and at $-1.75$ on L1,
inside the calibrated interval, which does not change. Over 20 periods the stable run at
$a=-1.90$ decays to about 1\% of its initial amplitude and the two halves of its window then
disagree, so it is indeterminate; over 8 to 12 periods it is stable ($\gamma=-0.032$ to
$-0.033$).

\subsection{Mechanism of the restabilisation branch}
\label{sec:fold}
\begin{figure}[t]
\centering
\includegraphics[width=\textwidth]{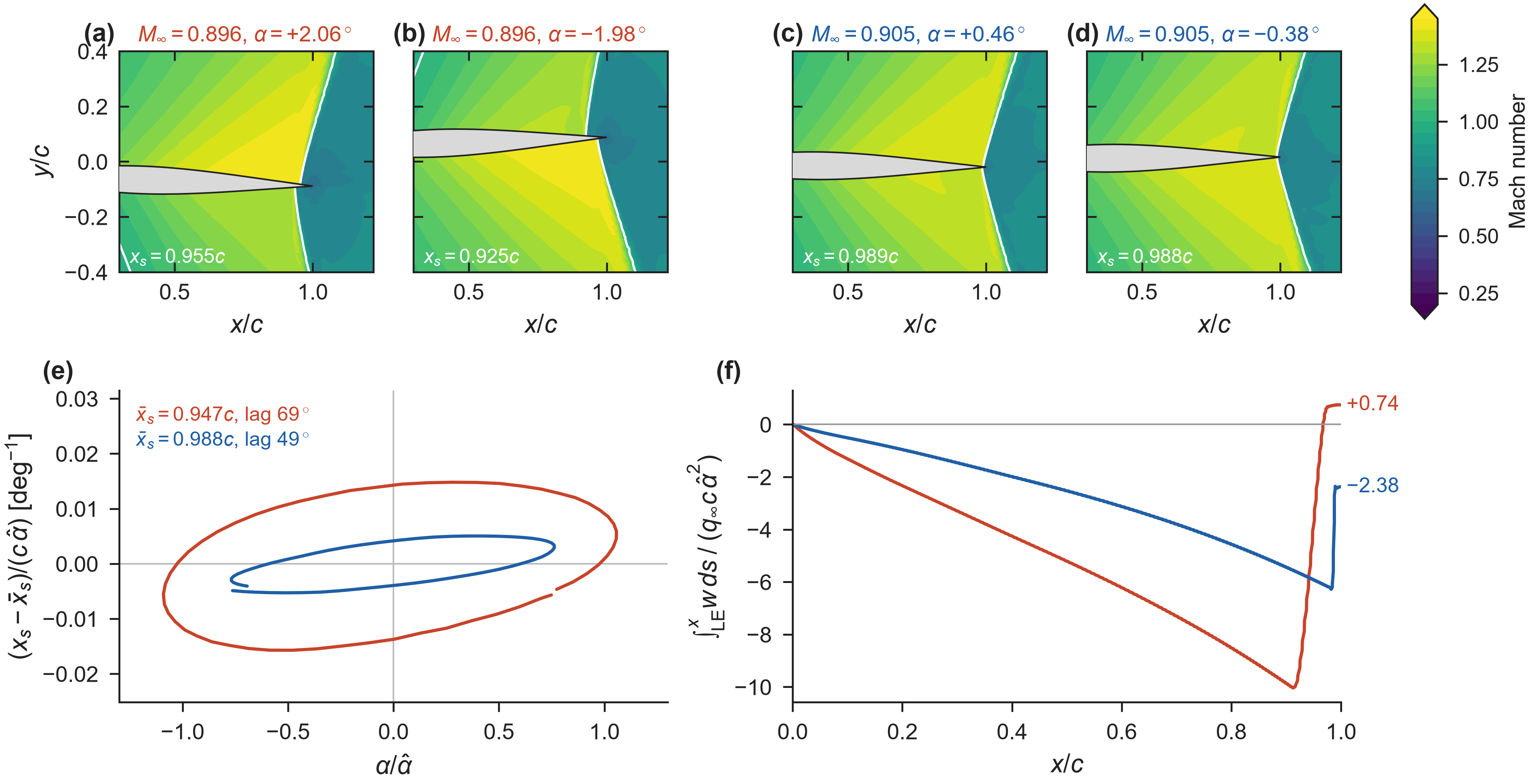}
\caption{Restabilisation on the upper branch: $\Vs=1.2$, $a=-2$, mesh L1, at $M_\infty=0.896$
(unstable, red) and $0.905$ (stable, blue). (a--d) Mach number at the nose-up and nose-down
extremes of a cycle (for the decaying stable run, its largest cycle after start-up), with
the upper-surface shock position. (e) Shock position against pitch over the last cycle,
normalised by the pitch amplitude, with the mean position and lag. (f) Running integral
from the leading edge of the work per cycle on both surfaces; end values: net work.}
\label{fig:fold}
\end{figure}

Two simulations on mesh L1 at $\Vs=1.2$ bracket the upper branch, which the learned map
places at $M_\infty=0.901$: $M_\infty=0.896$ is unstable
($\gamma=+0.009$) and $0.905$ stable ($-0.026$). The growing record is analysed with the poles of the four-cycle window,
and the three evaluations of the net work again agree, within 6\% ($+0.74$ and $-2.38$
from the surface, $+0.79$ and $-2.37$ from the loads, $+0.78$ and $-2.37$ from the growth
rate). The flow fields (\cref{fig:fold}a--d) show the difference. At $M_\infty=0.896$ each
supersonic region ends in a shock at $0.947c$ that moves by $0.060c$ over the cycle and lags
the pitch by $69^\circ$; at $0.905$ it extends to the trailing edge, where the shock sits at
$0.988c$ and moves by $0.0015c$. Reduced frequency ($k=0.125$ and 0.123) and mode (plunge to
pitch 1.97 and 1.98) barely differ. Per cycle the shock band ($0.90\le x/c\le1$) supplies
$+10.6$ and $+3.1$, while the loss upstream of it falls from
9.9 to 5.4 (\cref{fig:fold}f). Here the band reaches the trailing
edge, so no smooth field can be interpolated across it, and its integral contains all the
pressure work aft of $0.90c$; moving its lower limit between $0.85c$ and $0.92c$ changes the
fractions below by at most three points (\cref{tab:shockband}).

That a shock reaching the trailing edge stabilises the section is an established
explanation: \citet{isogai1981} found the flutter speed of case~A six times higher once the
shock had reached the trailing edge, shocks there have been argued to lose their mobility
and with it their destabilising effect \cite{bendiksen1992,bendiksen2011} or to move little
\cite{djayapertapa2004}, morphing that moves the shock to the trailing edge stabilises
earlier \cite{he2021}, and viscosity loses its influence where the shock sits near the
trailing edge \cite{karnick2017}.
The budget adds the size of each part at matched conditions: the shock excursion falls by
a factor of 40 and its input by 71\%, but the upstream damping falls by
45\% at the same time and offsets 59\% of the lost input; the section restabilises only because the shock
input falls further. On the lower branch the shock motion is the same on both sides and
the reduced frequency decides; on the upper branch the shock motion itself disappears.

\subsection{Why the fold closes}
\label{sec:closure}
\begin{figure}[t]
\centering
\includegraphics[width=\textwidth]{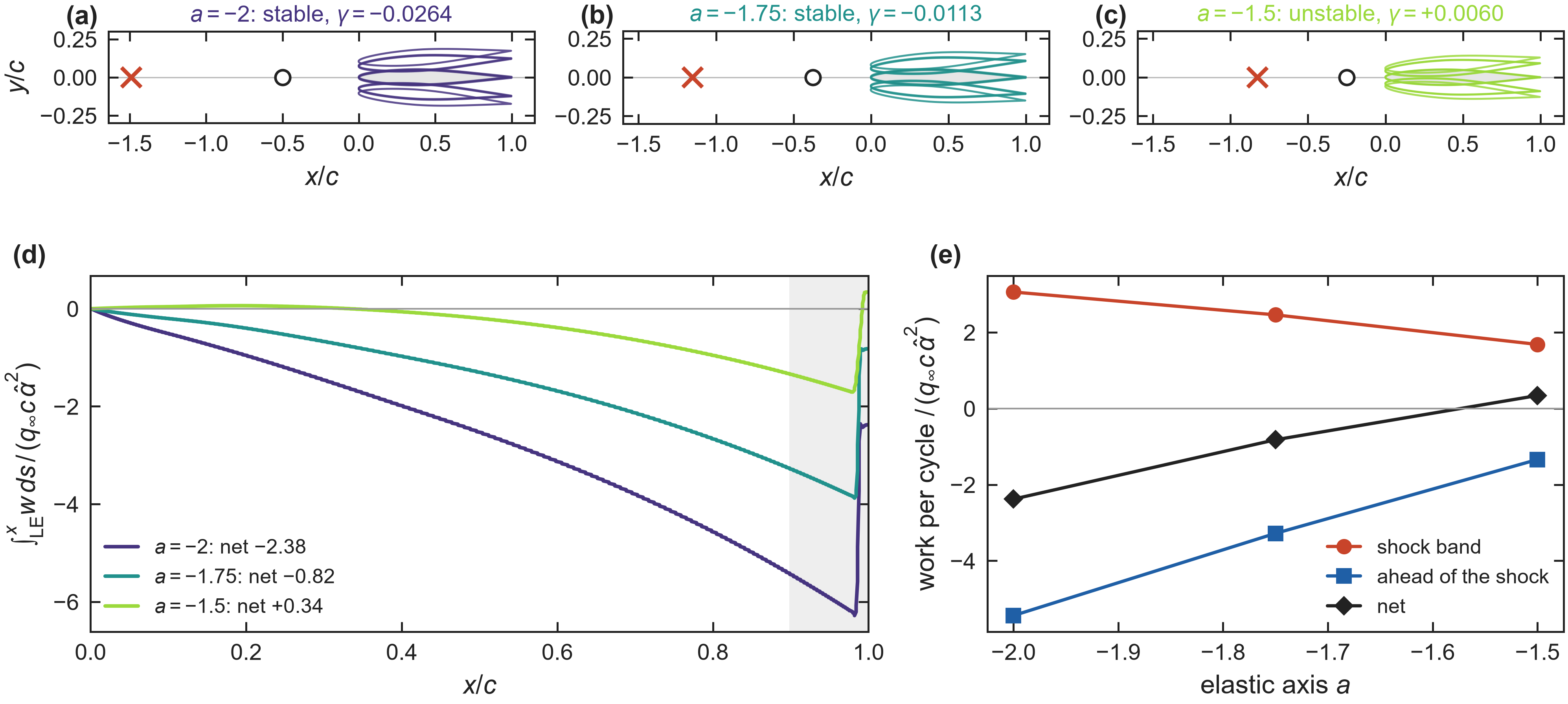}
\caption{Why the fold closes: one flow, three elastic axes ($M_\infty=0.905$, $\Vs=1.2$, mesh
L1, shock at the trailing edge). (a--c) Identified flutter mode, drawn at a pitch amplitude
of $4^\circ$ at eight phases, with the elastic axis (circle) and the pivot of the mode
(cross); titles give the damping over the last four cycles. (d) Running integral of the
work per cycle from the leading edge, both surfaces. (e) Work per cycle of the band swept by
the shock ($0.90\le x/c\le1$), of the surface ahead of it and in total, against the elastic
axis.}
\label{fig:closure}
\end{figure}

The restabilisation branch, and with it the fold, disappears as the elastic axis moves aft.
To see why, the restabilised condition of case~A, $M_\infty=0.905$, $\Vs=1.2$, was repeated
on mesh L1 with the elastic axis at $a=-1.75$ and $-1.5$ (\cref{fig:closure}). The flow is
the same in the three runs: the shock sits at the trailing edge ($x_s=0.988c$ for $a=-2$ and
$-1.75$, 0.989 on average for $-1.5$) and moves by at most $0.013c$, and the reduced
frequency is 0.123--0.124. The section is stable at $a=-2$ ($\gamma=-0.026$) and $-1.75$
($-0.011$) and unstable at $-1.5$ ($+0.020$ over the identification window, set by a
weak pole of nearly the same frequency with 5\% of the energy; the dominant pole gives
$+0.008$, four-period windows $+0.006$ to $+0.008$, and $+0.006$ over
the last four cycles, at $1.6^\circ$ of pitch). The net work per cycle again closes three
ways, within 2.2\% ($-2.38$, $-0.82$ and $+0.34$ from the surface; $-2.37$, $-0.81$ and $+0.35$
from the loads; $-2.37$, $-0.81$ and $+0.34$ from the growth rate).

What changes is the flutter mode. The plunge-to-pitch ratio falls from 1.98 to 1.55 and
1.15, with the plunge within $3.3^\circ$ of the pitch phase, so the pivot of the mode moves
from $1.49c$ to $1.15c$ and $0.83c$ ahead of the leading edge (\cref{fig:closure}a--c). For a
given pitch amplitude the surface then moves less, by a third at mid-chord and by a quarter
at the trailing edge between $a=-2$ and $-1.5$. The quasi-steady damping of the surface
ahead of the shock falls from 5.44 to 3.28 and 1.34 per cycle ($-75\%$), faster than the
input of the pinned shock, 3.06, 2.46 and 1.68 ($-45\%$); for lower band limits from
$0.85c$ to $0.94c$ these falls are $-74$ to $-77\%$ and $-43$ to $-46\%$, and the net changes
sign between $a=-1.75$ and $-1.5$ (\cref{fig:closure}e). At this condition the section
therefore restabilises only while the pivot of its flutter mode lies more than about one
chord ahead of the leading edge, which is why, in the family, the fold of case~A needs Isogai's far-forward
elastic axis.

In the family the centre of gravity moves with the axis (\cref{sec:formulation}). To
see whether the closure belongs to the axis or to the moving mass, the section was also
run with the centre of gravity fixed at its case~A position, $0.4c$ from the leading edge,
and the radius of gyration about it fixed at $0.49b$, so that $x_\alpha=-0.2-a$ and
$r_\alpha^2=0.24+x_\alpha^2$ change with $a$. The results are in \cref{tab:closurecheck}b.
The lower branch is unchanged: at $M_\infty=0.89$, $\Vs=1.2$ the section flutters for every
axis from $a=-1.75$ to $-1.3$ ($\gamma=+0.039$ to $+0.042$). The restabilisation, however,
extends further aft. At the condition of \cref{fig:closure}, on mesh L1, the section is
stable at $a=-1.75$ ($\gamma=-0.017$) and still at $-1.5$ ($-0.006$), where with the case~A
structure it flutters; at $M_\infty=0.89$, $\Vs=2.0$ and $a=-1.75$ its flutter mode is damped
with $\gamma=-0.10$, against $+0.002$ for the case~A structure. The pivot argument accounts
for this. With the centre of gravity fixed, the pivot of the mode moves less, from $1.23c$
to $0.97c$ ahead of the leading edge between $a=-1.75$ and $-1.5$, against $1.15c$ to
$0.83c$; the upstream damping falls by 30\% ($-3.96$ to $-2.75$) instead of 59\%, the shock
input by 11\% (2.71 to 2.41) instead of 32\%, and the net work stays negative ($-1.24$ and
$-0.34$, closed three ways within 5\%). The shock again sits at the trailing edge
($0.988c$), and the reduced frequency rises slightly, to 0.127 and 0.133. The closure of the
family is therefore not a property of the axis alone: it follows the pivot of the flutter
mode, which the centre of gravity, moving aft with the axis, helps to bring towards the
section.

Fixing the centre of gravity also brings in a second mode. At $\Vs=2.0$ and
$a\ge-1.6$ the section grows within about two pitch periods into an oscillation of
11--14$^\circ$ of pitch at 3.3--$4.0\,\omega_\alpha$. This is the second structural mode. In
vacuo it rotates about a point 0.08--$0.10b$ aft of the centre of gravity (in these runs
0.3--$0.5b$ ahead of it) and lies at $5.3\,\omega_\alpha$ for every member of the family; with
the centre of gravity fixed its frequency falls from 4.6 to $3.4\,\omega_\alpha$ as the axis
approaches the centre of gravity. At 128 steps per pitch period it is resolved by only
32--39 steps per cycle, too few to rely on its damping (\cref{sec:failures}), so the runs at
$a=-1.75$ and $-1.6$ were repeated with 256 steps (64--67 steps per cycle). The result does
not change. At $a=-1.6$ the pitch amplitude differs by at most 6\% during the growth, reaches
$10^\circ$ after 2.1 periods and settles at $11.2^\circ$ and $4.0\,\omega_\alpha$ with either step; at
$a=-1.75$ the section stays stable ($\gamma=-0.017$, against $-0.014$). The growth of the
second mode is therefore not an artefact of the time step, although it has been computed on
mesh L0 only. With the case~A structure no pole other than the flutter mode carries 5\% of
the energy in the runs of \cref{tab:closurecheck}a.

\section{Discussion}
\label{sec:discussion}
\emph{What a damping label contains.} A time-domain flutter label belongs to the simulation
protocol as well as to the configuration. At the bottom of the dip the half-degree mean
incidence left by the usual $1^\circ$ start outweighs every numerical setting, followed by
the mesh and a coarse time step (\cref{tab:sens}), and none of them shows in a converged
residual history. Benchmark
computations should therefore report the freestream incidence and the initial
condition of the unsteady run, the mean incidence that
results, including the
static deflection, and the number of time steps per period of the critical mode.

\emph{Why all of these are of order one per cent.} Near the boundary the net work per cycle
is 3--6\% of the shock input and changes sign over $\Delta\Vs=0.03$ (slope about 36 per unit
$\Vs$). A relative error $\epsilon$ in the shock work therefore shifts the boundary by about
$12.5\,\epsilon/36\approx0.35\,\epsilon$ in $\Vs$, or 0.65\% of $\Vf$ per per cent: anything
that changes the shock motion by a per cent, whether its grid resolution, the phase error
of a coarse step or the asymmetric shocks of a mean incidence, moves the flutter speed by a
comparable fraction. The coarse step stabilises even though it lowers the frequency
slightly (0.7--0.8\% with 32 steps), which alone would weaken the quasi-steady loss, so its
error must enter through the shock response or the time integration; we have not separated
the two.

\emph{Limitations.} The flow is inviscid; viscosity shifts the phase of the shock motion
and removes the multiple flutter points of case~A \cite{karnick2017}, so the fold mapped
here, and with it the critical elastic axis at which it closes, belong to the
Euler model; whether a critical axis exists in viscous flow, and where, remains to be
shown. The section is two-dimensional and only the elastic axis
is varied, with the structural parameters about it fixed. The conformal guarantee is
marginal, over queries distributed like the
calibration set, not conditional at every boundary point, and it is not a joint
statement about the topology of the boundary, and the stationary surrogate
gives its widest band where the boundary folds: in case~A the band reaches the bottom of
the $\Vs$ range between $M_\infty=0.868$ and 0.892, and in the family it leaves the closure
undecided for $-1.75\le a\le-1.70$. These are the places where further simulations would pay
most.

\section{Conclusions}
\label{sec:conclusions}
\begin{enumerate}
\item A time-domain damping label can be wrong without any residual warning: through
decoupled mesh motion, identification on a saturated limit cycle, or a coarse time step.
Each has a cheap check. The mean incidence of the oscillation is part of the definition of a
benchmark boundary: the half degree left by the common one-degree start raises $\Vf$ at the
dip by 2.8\%, more than any numerical setting, and is consistent
with the difference between two published values.
\item Conformal calibration on independent random simulations gives the learned boundary
stability calls with a distribution-free bound on their marginal error rate. For
case~A the coverage held at every level tested, at most 1.8\% of the calls made on held-out runs were wrong, and active
learning labelled 63 of 64 independent runs correctly
with 64 to 80 simulations, while random sampling
left 5--8\% of held-out runs wrongly labelled at 24 to 56.
\item In the Euler model, with the structural parameters about the axis fixed,
the fold of case~A is a property of a far-forward elastic axis. It exists
only for axes more than about 0.4 chord ahead of the leading edge and closes between 0.39
and 0.33 chord ahead, on two meshes; the bottom of the dip falls from $\Vs=0.62$ to the edge of the
domain as the axis moves from 0.75 chord ahead to the leading edge.
\item The two branches have different mechanisms. On the lower branch the shock motion and
the mode are unchanged and the falling reduced frequency lets the shock input overtake the
quasi-steady damping; it raises the one and lowers the other by similar amounts, so it
is the controlling parameter rather than a cause acting through the shock alone. On the upper branch the shock is pinned at the trailing edge and its
input collapses; the section restabilises only while the pivot of its flutter mode lies more
than about one chord ahead of the leading edge, because the damping of the supersonic
surface falls faster than the residual shock input as the pivot approaches. This is why the
fold closes. With the centre of gravity fixed instead, the pivot moves less, and at the
same flow condition the section is still restabilised with the axis 0.25 chord ahead of the
leading edge.
\item The flow is inviscid and two-dimensional. Viscosity removes the multiple flutter points
of case~A, so the natural next steps are Reynolds-averaged simulations of the same family and
a swept wing, where the same calibration and energy budgets apply unchanged.
\end{enumerate}

\appendix
\section{Solver settings and sensitivity studies}
\label{app:verification}

\subsection{SU2 settings of the coupled model}
\label{app:su2}
\Cref{tab:su2} lists the keys of the SU2 8.5.0 configuration that define the coupled
typical-section model and the mesh motion of the production runs; every configuration file
is in the code archive. Two keys need care. \texttt{REF\_ORIGIN\_MOMENT\_X} sets both the
moment reference and the pitch axis of the built-in model; without it the axis falls on the
coordinate origin, here the leading edge ($a=-1$). And the mesh must follow the aerofoil
through the legacy volumetric movement: with \texttt{DEFORM\_MESH=YES} and
\texttt{MARKER\_DEFORM\_MESH=(airfoil)} the deformation passes through the newer mesh
solver, and the surface barely follows the reported motion (\cref{sec:failures}).

The script \texttt{scripts/coupling\_check.py} of the archive reproduces
\cref{fig:failures}a,b. It runs both configurations for 64 time steps, half a pitch period,
at $M_\infty=0.85$, $a=-2$, $\Vs=1.7$ and $1^\circ$ freestream incidence with the surface
solution written at every step, takes the rotation of the chord line between the
leading-edge and trailing-edge nodes of each surface file, and tabulates it with the pitch
reported in the history file and the lift coefficient. With the volumetric movement the two
angles agree to $0.001^\circ$; with \texttt{DEFORM\_MESH=YES} the reported pitch reaches
$-7.28^\circ$ while the chord line rotates by at most $0.22^\circ$.

\begin{table}[t]
\centering
\caption{Keys of the SU2 configuration of the production runs (mesh L0; chord on
$0\le x\le1$).}
\label{tab:su2}
\small
\begin{tabular}{@{}>{\rr}p{60mm}>{\rr}p{44mm}>{\rr}p{46mm}@{}}
\toprule
Key & Value & Meaning \\
\midrule
\texttt{SOLVER} & \texttt{EULER} & \\
\texttt{CONV\_NUM\_METHOD\_FLOW}, \texttt{JST\_SENSOR\_COEFF} & \texttt{JST}; (0.5, 0.02) & \\
\texttt{NUM\_METHOD\_GRAD} & \texttt{GREEN\_GAUSS} & \\
\texttt{TIME\_MARCHING} & \texttt{DUAL\_TIME\_STEPPING-\allowbreak 2ND\_ORDER} & \\
\texttt{TIME\_STEP} & $2\pi/(128\,\omega_\alpha)=4.909\times10^{-4}$~s & 128 steps per pitch period \\
\texttt{INNER\_ITER} & 80 & \\
\texttt{TIME\_DISCRE\_FLOW}, \texttt{CFL\_NUMBER} & \texttt{EULER\_IMPLICIT}; 50 & pseudo-time iterations \\
\texttt{LINEAR\_SOLVER}, \texttt{\_PREC}, \texttt{\_ITER} & \texttt{FGMRES}; \texttt{ILU}; 10 & \\
\texttt{CONV\_FIELD}, \texttt{CONV\_RESIDUAL\_MINVAL} & \texttt{RMS\_DENSITY}; $-6$ & $-6.6$ on L1, $-7.2$ on L2 \\
\texttt{SURFACE\_MOVEMENT}, \texttt{MARKER\_MOVING} & \texttt{AEROELASTIC}; (\texttt{airfoil}) & built-in typical-section model \\
\texttt{AEROELASTIC\_ITER} & 5 & structure updated every five inner iterations \\
\texttt{REF\_ORIGIN\_MOMENT\_X} & $(1+a)/2$; $-0.5$ for case~A & elastic axis: pitch axis and moment reference \\
\texttt{FLUTTER\_SPEED\_INDEX} & $\Vs$ & $U_\infty=\Vs b\,\omega_\alpha\sqrt\mu$; the temperature keeps $M_\infty$ \\
\texttt{PLUNGE\_NATURAL\_FREQUENCY}, \texttt{PITCH\_NATURAL\_FREQUENCY} & 100; 100 & $\omega_h$, $\omega_\alpha$ [rad\,s$^{-1}$] \\
\texttt{AIRFOIL\_MASS\_RATIO} & 60 & $\mu$ \\
\texttt{CG\_LOCATION} & 1.8 & $x_\alpha$, semichords aft of the axis \\
\texttt{RADIUS\_GYRATION\_SQUARED} & 3.48 & $r_\alpha^2$, about the axis \\
\texttt{DEFORM\_STIFFNESS\_TYPE} & \texttt{INVERSE\_VOLUME} & volumetric movement \\
\texttt{DEFORM\_LINEAR\_SOLVER}, \texttt{\_PREC}, \texttt{\_ITER}, \texttt{\_ERROR} &
\texttt{CONJUGATE\_GRADIENT}; \texttt{ILU}; 1000; $10^{-9}$ & \\
\texttt{DEFORM\_NONLINEAR\_ITER} & 1 & \\
\texttt{DEFORM\_MESH}, \texttt{MARKER\_DEFORM\_MESH} & not set (\texttt{DEFORM\_MESH=NO}) &
\texttt{YES} decouples the mesh from the aerofoil \\
\texttt{AOA} & 0 & freestream incidence \\
\texttt{RESTART\_SOL}, \texttt{RESTART\_ITER} & \texttt{YES}; 2 & release from the steady $1^\circ$ solution \\
\bottomrule
\end{tabular}
\end{table}

\subsection{Sensitivity of the identified damping}
\label{app:ident}
\Cref{tab:ident} repeats the identification of \cref{sec:ident} with one setting changed
at a time: the energy threshold of the mode selection (1, 2, 5, 10, 20 and 50\%), the
discarded start (one to four pitch periods), the pencil order (6, 8, 10 and 12), the
channels (pitch and plunge together, or either alone) and the record length (8, 10 and 12
periods); a window of four pitch periods is also slid through the record. The runs that
bracket the bottom of the dip, and those clearly stable or unstable, keep their label in
every variant. Two runs near the closure of the fold are within 0.003 of neutral and change
label.

\begin{figure}[t]
\centering
\includegraphics[width=\textwidth]{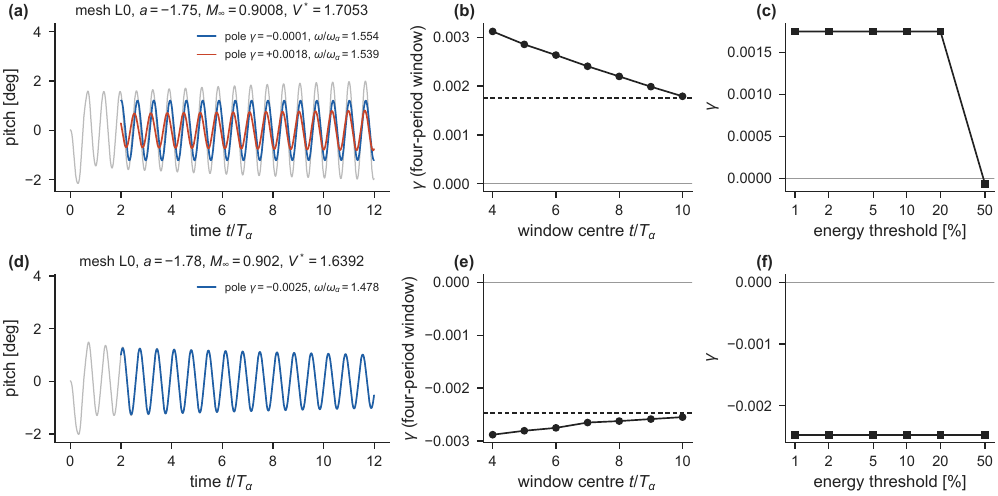}
\caption{Identification near the closure of the fold, mesh L0. Top: $a=-1.75$,
$M_\infty=0.9008$, $\Vs=1.7053$; bottom: $a=-1.78$, $M_\infty=0.902$, $\Vs=1.6392$. (a, d)
Pitch history (grey) and the contribution of each pole kept by the production fit over the
identification window; (b, e) damping coefficient in a four-period window against the
window centre, with the full-window value (dashed); (c, f) damping coefficient of the
production window against the energy threshold of the mode selection.}
\label{fig:identsens}
\end{figure}

\Cref{fig:identsens} shows the modal components of two neighbouring runs near the
closure. At $a=-1.78$ a single pole carries the response, and its damping is negative in
every window and for every threshold. At $a=-1.75$ two poles of nearly equal frequency,
$1.554$ and $1.539\,\omega_\alpha$, share the window: the stronger is neutral
($\gamma=-0.0001$), the weaker, with 36\% of its energy, grows ($+0.0018$), and the growth in
the sliding windows falls along the record towards the full-window value. The two
frequencies are $0.015\,\omega_\alpha$ apart, seven times closer than the Fourier resolution
of a ten-period window, $0.1\,\omega_\alpha$; only the parametric fit separates them, and
whether they are two modes or one mode whose damping drifts with amplitude cannot be decided
from this record. Such a run is therefore near-neutral, not confidently unstable, which is
how it enters the discussion of \cref{sec:family}. The upper-branch run at $a=-1.5$ also has
a weak second pole ($5\%$ of the energy) that sets $\gamma$ with the production threshold,
but its dominant pole and every window are unstable as well.

\begin{table}[t]
\centering
\caption{Damping coefficient ($\gamma\times10^{3}$) with the production identification,
its range over the variants of \cref{app:ident}, and its range over sliding four-period
windows; S: stable, U: unstable, I: indeterminate.}
\label{tab:ident}
\small\setlength{\tabcolsep}{5pt}
\begin{tabular}{@{}llrcrcr@{}}
\toprule
Run & Mesh & Production & Label & Variants & Labels & Windows \\
\midrule
\multicolumn{7}{@{}l}{Bottom of the dip: $M_\infty=0.85$, $a=-2$} \\
\quad $\Vs=0.53$ & L0 & $+0.88$ & U & $+0.87$ to $+0.98$ & U & $+0.85$ to $+0.99$ \\
\quad $\Vs=0.50$ & L1 & $-2.24$ & S & $-2.29$ to $-2.11$ & S & $-2.28$ to $-1.95$ \\
\quad $\Vs=0.53$ & L1 & $+1.46$ & U & $+1.42$ to $+1.67$ & U & $+1.38$ to $+1.72$ \\
\multicolumn{7}{@{}l}{Closure of the fold: $a$, $M_\infty$, $\Vs$} \\
\quad $-1.825$, 0.8984, 1.7605 & L0 & $-3.71$ & S & $-3.71$ to $-1.28$ & S & $-3.78$ to $-3.21$ \\
\quad $-1.78$, 0.902, 1.6392 & L0 & $-2.47$ & S & $-2.81$ to $-2.43$ & S & $-2.88$ to $-2.55$ \\
\quad $-1.75$, 0.9008, 1.7053 & L0 & $+1.75$ & U & $-0.07$ to $+2.77$ & U, I & $+1.79$ to $+3.12$ \\
\quad $-1.75$, 0.89, 2.0 & L1 & $+0.18$ & I & $-0.65$ to $+2.88$ & I, U & $+0.91$ to $+2.42$ \\
\multicolumn{7}{@{}l}{Upper branch: $M_\infty=0.905$, $\Vs=1.2$} \\
\quad $a=-1.75$ & L1 & $-11.38$ & S & $-11.92$ to $-11.21$ & S & $-11.96$ to $-11.34$ \\
\quad $a=-1.5$ & L1 & $+20.01$ & U & $+6.77$ to $+20.01$ & U & $+6.15$ to $+7.71$ \\
\bottomrule
\end{tabular}
\end{table}

\subsection{Verification runs at the closure}
\label{app:closure}
\Cref{tab:closurecheck} lists the runs that verify the closure of the fold
(\cref{sec:family,sec:closure}): with the case~A structure about the axis on two meshes,
with two time steps and for four record lengths, and with the centre of gravity fixed.

\begin{table}[t]
\centering
\caption{Runs at the closure of the fold, zero mean incidence: damping coefficient
$\gamma\times10^{3}$ and label (S stable, U unstable, I indeterminate). (a) Case~A structure
about the axis, $M_\infty=0.89$; the runs at $a\ge-1.70$ reach the $3^\circ$ cap within eight
periods, so their values do not depend on the record length; $^\ast$runs of \cref{sec:family}.
(b) Centre of gravity fixed at $0.4c$, mesh L0 and 128 steps per period unless stated; LCO: oscillation of the second
mode, with its pitch amplitude; $\omega/\omega_\alpha$: frequency of the critical pole at
$\Vs=2.0$.}
\label{tab:closurecheck}
\small\setlength{\tabcolsep}{3.5pt}
\newcommand{\lb}[1]{\,\makebox[0.75em][l]{#1}}  %
\begin{tabular}{@{}lrrrrrrr@{}}
\toprule
\multicolumn{8}{@{}l}{(a) Case~A structure about the axis} \\
$a$ & $-1.90$\lb{} & $-1.80$\lb{} & $-1.75$\lb{} & $-1.70$\lb{} & $-1.65$\lb{} & $-1.60$\lb{} & $-1.50$\lb{} \\
\midrule
\multicolumn{8}{@{}l}{$\Vs=2.0$, L0, 128 steps per period} \\
\quad 8 periods  & $-32.0$\lb{S} & $-8.5$\lb{S} & $+4.3$\lb{U} & $+16.2$\lb{U} & $+7.7$\lb{U} & $+14.6$\lb{U} & $+16.1$\lb{U} \\
\quad 10 periods & $-32.8$\lb{S} & $-8.5$\lb{S} & $+3.2$\lb{U} & $+16.2$\lb{U} & $+7.7$\lb{U} & $+14.6$\lb{U} & $+16.1$\lb{U} \\
\quad 12 periods & $-33.5$\lb{S} & $-8.8$\lb{S} & $+2.3$\lb{U} & $+16.2$\lb{U} & $+7.7$\lb{U} & $+14.6$\lb{U} & $+16.1$\lb{U} \\
\quad 20 periods & $-34.9$\lb{I} & $-8.4$\lb{S} & $+1.3$\lb{U} & $+16.2$\lb{U} & $+7.7$\lb{U} & $+14.6$\lb{U} & $+16.1$\lb{U} \\
\multicolumn{8}{@{}l}{$\Vs=2.0$, L0, 256 steps per period} \\
\quad 8 periods  & & $-8.4$\lb{S} & $+4.3$\lb{U} & $+16.1$\lb{U} & & & \\
\quad 10 periods & & $-8.4$\lb{S} & $+3.2$\lb{U} & $+16.1$\lb{U} & & & \\
\quad 12 periods & & $-8.6$\lb{S} & $+2.3$\lb{U} & $+16.1$\lb{U} & & & \\
\multicolumn{8}{@{}l}{$\Vs=2.0$, L1, 128 steps per period} \\
\quad 12 periods & $-45.5$\lb{S$^\ast$} & $-19.2$\lb{S} & $+0.2$\lb{I$^\ast$} & $+17.6$\lb{U} & & $+16.3$\lb{U$^\ast$} & \\
\multicolumn{8}{@{}l}{$\Vs=1.2$, L0, 128 steps per period} \\
\quad 12 periods & & & $+52.0$\lb{U} & & & $+58.3$\lb{U} & \\
\midrule
\multicolumn{8}{@{}l}{(b) Centre of gravity fixed} \\
$a$ & $-1.75$\lb{} & $-1.60$\lb{} & $-1.50$\lb{} & $-1.30$\lb{} & & & \\
\midrule
$M_\infty=0.89$, $\Vs=1.2$ & $+39.3$\lb{U} & $+41.6$\lb{U} & $+41.6$\lb{U} & $+38.6$\lb{U} & & & \\
$M_\infty=0.89$, $\Vs=2.0$ & $-13.8$\lb{S} & LCO $11^\circ$\lb{} & LCO $12^\circ$\lb{} & LCO $14^\circ$\lb{} & & & \\
\quad $\omega/\omega_\alpha$ & 3.81\lb{} & 4.04\lb{} & 3.81\lb{} & 3.29\lb{} & & & \\
\quad 256 steps per period & $-16.5$\lb{S} & LCO $11^\circ$\lb{} & & & & & \\
\quad $\omega/\omega_\alpha$ & 3.84\lb{} & 4.01\lb{} & & & & & \\
$M_\infty=0.905$, $\Vs=1.2$, L1 & $-17.3$\lb{S} & & $-6.0$\lb{S} & & & & \\
\bottomrule
\end{tabular}
\end{table}

\subsection{Limits of the shock band}
\label{app:shock}
\Cref{tab:shockband} gives the regional work for several limits of the shock band. At
the bottom of the dip the band lies inside the chord, and the shock work is the band
integral less the smooth part interpolated across the band (\cref{sec:energy}). On the upper
branch the shock sits at $0.947c$ ($M_\infty=0.896$) or at the trailing edge ($0.905$), so the
band runs to the trailing edge and only its lower limit is varied; a limit of $0.94c$ would
cut through the shock of the unstable run and is used only for the closure runs.

\begin{table}[t]
\centering
\caption{Work per cycle done by the flow, normalised by $q_\infty c\,\hat\alpha^2$, for
several limits of the shock band (mesh L1). (a) Bottom of the dip, $M_\infty=0.85$, $a=-2$,
at $\Vs=0.50$ / 0.53; the share is that of the shock work in the change of the net work,
$+1.09$. (b) Upper branch, $\Vs=1.2$, $a=-2$, from $M_\infty=0.896$ to 0.905. (c) Closure,
$M_\infty=0.905$, $\Vs=1.2$, from $a=-2$ to $-1.5$.}
\label{tab:shockband}
\small
\begin{tabular}{@{}lllll@{}}
\toprule
\multicolumn{5}{@{}l}{(a) Bottom of the dip, $\Vs=0.50$ / 0.53} \\
Band $x/c$ & Band integral & Smooth part & Shock work & Share \\
\midrule
0.69--0.76 & 12.29 / 12.82 & $-1.42$ / $-1.38$ & 13.71 / 14.20 & 45\% \\
0.70--0.75 & 12.85 / 13.27 & $-1.06$ / $-0.93$ & 13.91 / 14.21 & 27\% \\
0.68--0.77 & 11.91 / 12.45 & $-1.78$ / $-1.70$ & 13.69 / 14.16 & 42\% \\
0.67--0.78 & 11.46 / 12.02 & $-2.16$ / $-2.09$ & 13.62 / 14.11 & 45\% \\
0.65--0.80 & 10.67 / 11.27 & $-2.82$ / $-2.70$ & 13.49 / 13.97 & 44\% \\
\midrule
\multicolumn{5}{@{}l}{(b) Upper branch, $M_\infty=0.896$ / 0.905} \\
Lower limit & Band & Ahead of the band & Fall of band, of ahead & Offset \\
\midrule
0.85 & 9.88 / 2.59 & $-9.13$ / $-4.97$ & 74\%, 46\% & 57\% \\
0.88 & 10.32 / 2.88 & $-9.58$ / $-5.26$ & 72\%, 45\% & 58\% \\
0.90 & 10.60 / 3.06 & $-9.86$ / $-5.44$ & 71\%, 45\% & 59\% \\
0.92 & 10.69 / 3.23 & $-9.95$ / $-5.61$ & 70\%, 44\% & 58\% \\
\midrule
\multicolumn{5}{@{}l}{(c) Closure, $a=-2$ / $-1.75$ / $-1.5$} \\
Lower limit & Band & Ahead of the band & Change of band, of ahead & \\
\midrule
0.85 & 2.59 / 2.12 / 1.47 & $-4.97$ / $-2.95$ / $-1.13$ & $-43\%$, $-77\%$ & \\
0.88 & 2.88 / 2.33 / 1.60 & $-5.26$ / $-3.15$ / $-1.26$ & $-44\%$, $-76\%$ & \\
0.90 & 3.06 / 2.46 / 1.68 & $-5.44$ / $-3.28$ / $-1.34$ & $-45\%$, $-75\%$ & \\
0.92 & 3.23 / 2.58 / 1.76 & $-5.61$ / $-3.40$ / $-1.42$ & $-45\%$, $-75\%$ & \\
0.94 & 3.42 / 2.72 / 1.85 & $-5.81$ / $-3.54$ / $-1.51$ & $-46\%$, $-74\%$ & \\
\bottomrule
\end{tabular}
\end{table}

\FloatBarrier
\section*{Informed consent}
Not applicable.

\section*{Author contributions}
All authors have accepted responsibility for the entire content of this manuscript and
approved its submission.

\section*{Conflict of interest}
The authors declare that they have no conflict of interest.

\section*{Data availability}
The code, solver configurations, meshes, run databases and figure scripts are available
at \url{https://github.com/i4mGr0ot/Active-Learning-Transonic-Flutter-Prediction} and archived
at \url{https://doi.org/10.5281/zenodo.23415511}.

\bibliographystyle{IEEEtranN_doi}
\bibliography{refs}

\end{document}